# Airport-Airline Coordination with Economic, Environmental and Social Considerations

I. Aasheesh Dixit (Corresponding Author), Institute of Management Technology, Dubai, 345 006, Dubai, UAE. aasheesh@imt.ac.ae

II. Patanjal Kumar, Indian Institute of Management Rohtak, Rohtak-124010 Haryana, India. patanjaliitiim@gmail.com

III. Suresh Kumar Jakhar, Indian Institute of Management Lucknow, Lucknow, 226 013, India. skj@iiml.ac.in

**Abstract:** In this paper, we examine the effect of various contracts between a socially concerned airport and an environmentally conscious airline regarding their profitability and channel coordination under two distinct settings. First, we consider no government interventions, while in the second, we explore government-imposed taxations to curb emissions. Furthermore, we investigate the impact of passenger greening sensitivity, greening cost, and consumer surplus coefficient on conveyance fees, ticket fare, greening level and the channel welfare. Our analysis shows that the revenue sharing and linear two-part tariff contracts coordinate the decentralised airport-airline channel. Our findings also reveal that players' greening and social efforts can improve both the welfare and efficiency of the channel simultaneously. Importantly, under government interventions, taxation does help improve the greening level of the channel in both coordinating and non-coordinating contracts. However, the greening level in the non-coordinating contracts with taxation is still less than the coordinating contracts even without tax. Finally, we also extended the model to include a duopoly airline market with pricing and greening competition. We analyze the effect of competetiton between airlines on airport utility, airline profit, ticket fare and greening level.

1.     **Introduction**

Airports generate revenue from two sources, namely: (i) aeronautical activities[1] and (ii) commercial activities[2]. The former is proportional to flight frequency and passenger traffic, while the latter depends only on passenger traffic (Chang et al., 2016; Zhang & Zhang, 1997). Therefore, increasing air traffic has led to a twofold rise in airport business, as its revenue has increased from both sources (Francis et al., 2004). However, even with growing demand, airlines are finding it difficult to maintain their profit margins with rising input costs and market competition. For example, the US airlines faced margin reductions of 5% to 6% in 2018, marking the third straight year of margin contraction. Unfortunately, the operating margins are expected to narrow further in the upcoming years (Stalnaker, 2019). Add to that the ever-increasing global warming concerns that has led policy makers to introduce additional charges, such as emission taxes, which have effectively forced airlines to increase investments in efficient and green technologies to reduce carbon emission (Masiol & Harrison, 2014). These investments have burdened the cost structure of the airlines, affecting their margins thereof.

Notably, even though air transportation industry today, do recognise the significance of social and environmental aspects, (as captured in Munich-Lufthansa agreement and Norway's Oslo Airport and its signatory airline agreement discussed below) increasing operating costs leaves less room for airlines to engage in social and ecological activities concurrently. Due to this, very few can achieve a sustainable growth including economic, environmental, and social elements altogether (Elkington, 1998). One of the major hindrances to attain sustainable growth is the lack of a coherent coordination strategy between industry agents (Ryder, 2014). For instance, in the aviation industry, an airline's performance determines the airport's demand, while the airport needs to facilitate the airline by providing the necessary infrastructure. This inter-dependency has to be well coordinated, failure of which would lead to a limited economic growth, environmental protection and social development.

---
[1] Aeronautical revenues are paid by airlines depending on the use of runways, terminals and gates, which are proportional to flight frequency.
[2] Commercial activity includes terminal concessions (duty-free shops, restaurants, etc.), car parking and rental, which are proportional to passenger traffic.

Nevertheless, the airline-airport coordination seems to have picked up in recent times under a few cases. For instance, the first glimpse of a contract incorporating the environmental aspect has been seen in the Munich-Lufthansa agreement. Under this agreement, terminal 2 of the airport has been exclusively devoted to handling Lufthansa's operations, which in turn has pledged to continue investing in a modern fleet with fuel-efficient aircrafts, aiming to bring its emission levels down by 25% (Noëth, 2019). Another example of vertical coordination is Norway's Oslo Airport and its signatory airline; the former encourages the airline to use sustainable aviation fuels for its operations by providing incentives. The use of biofuels by the airline has resulted in significant environmental benefits (Baxter, 2020).

However, incorporating all aspects of sustainable growth to achieve all-round development by using contracts, has altogether remained an unexplored area in the aviation sector. Based on the examples cited above, we note that there hasn't been sufficient literature that has focused on exploring the importance of 'coordination' for profitability, planet (environmental) and people (social) in the airline industry. Thus, in this paper, we examine a dyadic model with one airline-one airport market structure with a single origin-destination (OD) route and explore four different contract types. We assume that the airline takes environmental responsibility, and the airport is involved in corporate social responsibility (CSR). We also explore the effectiveness of government intervention under these four different contracts. Herein, it may be worth noting that with government intervention, the airline pays an emission tax, which in turn is used for the social welfare of the local community. Our motivation is to explore how airline-airport can support each other with the help of different contracts to achieve sustainable growth in the aviation sector. Primarily, we seek answers to the following questions:

(i) Under what circumstances can airport-airline coordination achieve better economic, environmental, and social results?

(ii) What is the effect of environmental and social considerations on factors such as airport charges, ticket fares, and demand for air travel?

(iii) How can the use of contracts enhance mutual support and eliminate the problem of incoordination?

(iv) What is the role of government intervention in achieving sustainable growth in the aviation sector?

(v) How does contract affect the duopoly airline market with greening and pricing competition to achieve better economic, environmental, and social results?

Notably, our study demonstrates that channel performance and its efficiency improve when players engage in coordinating contracts instead of solely working for their individual goals under non-coordinating contracts. Interestingly, players achieve a higher level of greening and social welfare. Moreover, our findings show that revenue sharing and liner two-part tariff contracts perfectly coordinate the airport-airline channel. Regarding government intervention, we observe that taxation does help improve the greening level; however, it is still less in non-coordinating contracts with government intervention than coordinating contracts even without tax. We also find that the airline profit decreases with higher taxation and higher greening investment costs. We further analyse the result of the duopoly airline market with pricing and greening competition and undertake sensitivity analysis to support our findings.

The remainder of the paper is organised as follows: Section 2 provides a brief review of channel coordination in the aviation sector. Section 3 proposes a model with airline and airport's objective functions without government intervention. Section 4 briefly explains the different contract types that are studied. Section 5 provides analytical results. Section 6 presents the numerical analysis of the equilibrium results. Section 7 extends the coordination model by considering government intervention, and Section 8 provides a summary and conclusions. Finally, Section 9 provides future research directions.

## 2. Literature review

This manuscript mainly discusses three aspects: coordination between airport and airline, environmental concer and welfare concern. Therefore, in this section, we present their relevant literature and list the key research questions. We also compare this study with extant literature in Table 1.

Before liberalisation, both the airlines and airports were owned and operated by governments in many countries across the globe. They were public entities, and the need to explore any possible coordination strategies between the two were ignored. However, with the increasing trend of deregulation, airports and airlines began to realise the need for vertical agreements (Fu et al., 2010). The studies of Fu & Zhang (2010) and Barbot (2009) found that cooperation between an airport and its airlines may be beneficial for both players in terms of increased traffic volume and operational efficiencies.

As airports around the world are being commercialised, Starkie (2001) claimed that due to an increasing focus on non-aeronautical revenue, airports are less likely to charge high aeronautical fees. However, Carney & Mew, (2003) seemed to differ; they argued that the monopolistic nature of the airport business, could possibly encourage the airports to charge an exorbitant aeronautical fee. Zhang & Zhang (2003) along with Oum et al. (2004) added that an increase in 'aeronautical fee' (or service charges, as they called it), might effectively reduce passenger footfall; hence, airports are apprehensive in increasing aeronautical charges. Herein importantly, both growth in passenger footfall and improvement in operational efficiency, may be achieved by improved airport-airline coordination (Auerbach & Koch, 2007).

Gillen & Morrison (2003) on the other hand, suggested that there could be revenue-sharing contracts between airport-airline, which may actually internalise non-aeronautical revenue and reduce the travelers' total travel costs thereof. Fu & Zhang (2010) studied the effects of revenue sharing; that is, an airport sharing a fraction of its non-aeronautical revenue with one or more airlines serving the airport. The airport offers the airline a 'take it or leave it' contract, sharing thereby a proportion of its concession revenue. Nevertheless, the study did not consider aeronautical revenue, like landing fee and airport improvement fee. Alfonso (2012) in his doctoral thesis, discussed the importance of 'incentives' in a coordination between airlines and airports. Specifically, he studied the airport pricing of aeronautical fee and non-aeronautical activities and confirm a positive externality of non-aeronautical activities on the aeronautical charge. The author concluded that the airports have incentives to reduce the aeronautical charge so as to increase passengers' dwell time and their consumption of concession goods. Czerny & Zhang (2015) studied revenues derived from non-aeronautical source and per-flight fees to find the optimal airport charges, with a condition of airport cost recovery. Choo (2014) showed that an increase in per-passenger charges is associated with a reduction of per-flight costs; however, he noted that there's a possibility of vertical coordination, but fell short of discussing its associated benefits. Yang et. al. (2015) analysed airport–airline vertical arrangements using Nash bargaining game. They concluded that when there's a vertical arrangement, the airlines would perform better if they have greater market power; airports would perform better if their charges are higher, or the airport itself cares about social welfare. Xiao et al. (2016) also did study the airport-airline vertical coordination; however, they used the airport's capacity as a decision variable under the condition of demand uncertainty. Thus, based on these few examples from extant literature, we note that while most of the studies have focused on profit and the social dimensions, no study has effectively considered all three aspects (i.e. economic, environmental and social) of triple bottom line (TBL) in the airline industry for a coordinated market setting. Moreover, we also did not find any study in extant literature that has even looked to incorporate environmental issues in analysing the coordinating contracts in the industry.

With an estimated 200% increase in the emission level, the environmental hazard of the airlines industry is a growing area of concern (Masiol & Harrison, 2014). Albeit limited, there has been an attempt to curb emissions in recent times (Caro et al., 2013). Green economists advocate the use of well-designed carbon tax to reduce the emission levels significantly (Lin & Lu, 2015). In fact, they vouch for policies that directly target environmental goals via a Pigouvian tax or permit policies, which incidentally, are a common way to manage environmental problems. However, Sheu & Li (2014) noted that carbon permits/pricing is a reactive measure, while greening 'investment' is a proactive measure. In this study, we choose to discuss two settings; the first, wherein both airport and airline enter into a contract to promote greening investment and social responsibility. The second, we include the 'government', which imposes a tax on the airline for its carbon emissions, and uses the collected tax to improve the welfare of the local community. At their end, the airlines could reduce their long-term carbon emission by greening investment (GI), while cutting tax cost. Notably, GI includes investments in innovation, R&D in searching for fuel alternatives, upgrading fleet, and improving technology (Dray et al., 2014). Several studies in this regard have suggested that airlines must be incentivised to switch from conventional fuels to biofuels (Lu, 2018). The exemplary contract between Lufthansa and the Munich airport may be considered as a pioneering act in this regard. However, it would certainly be interesting to look into the aspect of 'coordination', and how it could help airlines improve their greening level, while drive the aviation market towards sustainability. We extend upon extant literature on the airport–airline vertical cooperation considering economic, environmental, and social progress. In the extant literature the incorporation of all three aspects (economic, environmental, and social) together is also referred as triple bottom line (TBL). We develop a vertical airline-airport model to improve coordination by considering economic, environmental, and social parameters. Some of the key features of our model are:

(i) We consider a greening airline that invests in improving its greening levels. We try to explore the impact of greening costs and green-sensitive passengers on the profitability of the airline, utility of the airport and social welfare of the channel.

(ii) The welfare-maximising airport captures the people (social) aspect of the TBL, which contributes to consumer surplus attributed to corporate social responsibility (CSR).

(iii)     We explore four different contracts to identify the one that possibly best coordinates the dyad to maximise the objectives of both the airport and the airline.

(iv)     We extended the model to include government intervention, where the airline pays an emission tax. We analyse and compare the results with and without tax regime.

Importantly, to the best of our knowledge, this is the first study that has looked at the vertical airline industry, while incorporating economic, environmental, and social aspects altogether. Through our study, we show that airports can significantly reduce aeronautical charges, and also improve their utility. In fact, they could also help the airlines in increasing their greening level, which in turn would make up for a decrease in revenue from the reduced aeronautical charges, because of an increased passenger count that would hike up the non-aeronautical revenues. Interestingly, in the case of government levying tax on the airline, the greening level would increase, but the overall channel performance would significantly decrease.

Table 1: Comparison of literature

| Authors | Airport airline coordination | Environmental aspect | Welfare aspect | Methodology |
|---|---|---|---|---|
| Fu & Zhang (2010) | x | | x | Game theory |
| Barbot (2009) | x | | | Game theory |
| Starkie (2001) | x | | | Concept paper |
| Carney & Mew, (2003) | x | | | Concept paper |
| Zhang & Zhang (2003) | x | | x | Analytical |
| Oum et al. (2004) | x | | x | Game theory |
| Auerbach & Koch (2007) | x | | | Case studies |
| Gillen & Morrison (2003) | x | | | Location model |
| Fu & Zhang (2010) | x | | x | Game theory |
| Alfonso (2012) | x | | | Game theory |
| Czerny & Zhang (2015) | x | | | Game theory |
| Choo (2014) | | | | Empirical study |
| Yang et. al. (2015) | x | | | Empirical study |
| Xiao et al. (2016) | x | | | Game theory |
| Masiol & Harrison, 2014 | | x | | Literature review |
| Caro et al., 2013 | | x | | Game theory |
| Lin & Lu, 2015 | | x | | Scenario |

| Sheu & Li (2014) | x | x |   | Game theory |
| --- | --- | --- | --- | --- |
| Dray et al., 2014 |   | x |   | Empirical study |
| Lu, 2018 |   | x |   | Cost benefit |
| Current study | x | x | x | Game theory |

## 3. Model: Without government intervention

We build a general vertical structure with one airline and one airport. We start by describing a market with an environmentally conscious airline and CSR-oriented airport. The list of notations used in subsequent sections is provided in Table 2.

**Table 2: Notations**

| S. No. | Term | Symbol |
| --- | --- | --- |
| 1 | Market potential | $\alpha$ |
| 2 | Passenger's sensitivity to ticket fare | $\beta$ |
| 3 | Ticket fare | $p$ |
| 4 | Demand of airline | $q$ |
| 5 | Greening cost per unit | $I$ |
| 6 | Frequency of airline | $f$ |
| 7 | Schedule-delay parameter | $\gamma$ |
| 8 | Passenger's sensitivity to greening | $\xi$ |
| 9 | Level of greening | $\theta$ |
| 10 | Desired level of greening | $\theta_0$ |
| 11 | Tax level | $t$ |
| 12 | Operating cost of airport | $c_{AP}$ |
| 13 | Operating cost of airline | $c_{AL}$ |
| 14 | Conveyance fee per passenger | $w$ |
| 15 | Non aeronautical revenue | $w'$ |
| 16 | Fraction of consumer surplus | $\mu$ |
| 17 | Landing charge per flight | $c$ |
| 18 | Revenue-sharing fraction of airline | $\psi$ |
| 19 | Cost-sharing fraction of airport | $\phi$ |
| 20 | Lump sum tariff paid by airline to airport | $L$ |
| 21 | Profit | $\pi$ |
| 22 | Reservation profit of airline | $\bar{\pi}^{AL}$ |
| 23 | Utility | $U$ |
| 24 | Social welfare efficiency | $SWE$ |
| 25 | Greening efficiency | $GE$ |
| Superscript |   |   |
| 26 | Airport | $AP$ |
| 27 | Airline | $AL$ |
| Subscript |   |   |

| 28 | Centralized setting | $CENT$ |
|---|---|---|
| 29 | Decentralised setting | $D-CENT$ |
| 30 | Cost-sharing contract | $CSC$ |
| 31 | Revenue-sharing contract | $RSC$ |
| 32 | Linear two-part tariff contract | $LTT$ |
| 33 | With tax case | $T$ |

**3.1 Demand function**: We consider demand ($q$) to be a function of fare ($p$), frequency ($f$) and the greening level ($\theta$). Passengers generally prefer airlines with higher flight frequency, as it helps in reducing schedule delays[3]. The higher greening level helps the airline to project an ecologically responsible image. Therefore, environmentally sensitive passengers rank such airlines higher in terms of service quality (Laroche et al., 2001), and have the propensity to pay higher for such services (Yaacob & Zakaria, 2011). The demand function is downward sloping in ticket fare and upward sloping in frequency and greening. This assumption is similar to Brueckner & Girvin (2008) and Ghosh & Shah (2012). Thereby, the proposed demand equation is a function of the market potential ($\alpha$), passenger's sensitivity to ticket fare ($\beta$), schedule-delay parameter ($\gamma$), passenger's sensitivity to greening ($\xi$), fare ($p$), frequency ($f$) and the greening level ($\theta$), as shown below

$$q = \alpha - \beta p - \frac{\gamma}{f} + \xi\theta \qquad (1)$$

**3.2 Cost structure.** We divide the airline cost structure into three parts:

(i) Operating cost (OC). We assume that the airline incurs an average per-passenger operating cost ($c_{AL}$), which is equal for all passengers. Hence, the total OC is denoted by $q.c_{AL}$;

(ii) Greening Investment (GI) cost: As the airline markets its services as being environmentally friendly, it incurs associated higher GI cost to reduce energy consumption and emission. For example, the use of bio-fuel reduces emission levels, but are 2.5-6 times expensive than conventional fuel (IEA, 2019). The extra costs for switching from traditional fuel to biofuel is equivalent to 41% of the average one-way economy ticket fares (Lu, 2018). In case, the airline leases a fuel-efficient aircraft, the payback period varies from 5-9 years, depending on economic conditions (Tsai et. al., 2012). Though the greening initiative has an inherent cost, they are expected to reduce the emission levels, and thereby make the airlines more environment-friendly.

---

[3] 'Scheduled Delay' is defined as the difference between the passenger's preferred departure time and nearest flight time and is inversely proportional to the airline's flight frequency $f$ (Richard, 2003).

The cost of green investment is assumed to be non-linear function, given by $I.\theta^2$, where $I$ is an investment cost parameter and $\theta$ is the greening level that airline achieve. The cost is increasing and quadratic in nature with convex property, which reflects the idea that greening firms often tend to capture the 'low hanging fruits' first with subsequent improvements becoming more difficult (Walley & Whitehead, 1994). Further, the return from R&D expenditures or large fleet investment is best captured by the convex costs attributed to diminishing returns. The case aptly fits into our model as green improvement is achieved through large R&D based investments or fleet replacement (Bhaskaran & Krishnan, 2009).

(iii) Airport charges (AC). Airport charges are the revenue earned by the airport in exchange for services provided. We consider the two most important charges:

a. Landing fees: The airline pays a fixed charge per flight and the total cost is given by $c.f$ and

b. Conveyance fees (CF): Airports charge a conveyance fee per passenger to cover for the airport infrastructure development. These have become an important revenue source for airports worldwide, as they derive revenue equal to the landing fees (Zhang & Czerny, 2012). For instance, the construction cost for Hong Kong airport's third runway is partly covered by an increase in the per-passenger based airport charges. We define total CF as $(w.q)$, where $w$ is $CF$ per passenger. Hence, the total airport charges are given by the sum of landing fees and conveyance fees, $c.f + w.q$

Finally, the profit function $(\pi)$ of the greening airline is defined as the difference between revenue $(p.q)$ and the total cost incurred;

$$Profit\ (\pi^{AL}) = (p.q) - (I.\theta^2 + c.f + w.q + q.c_{AL}) \qquad (2)$$

The airport's revenues come from two sources: aeronautical revenue (CF and landing charges) and non-aeronautical revenues (from duty-free shops, car hire/parking, rental/retail shops and restaurants) (Bitzan & James, 2017). Aeronautical revenue is the sum of per flight charges and conveyance fee per passenger. While non-aeronautical revenue is the average revenue generated per passenger by using extra services of the airport, and is given by $w'$. Thus, the total revenue of the airport is given by $c.f + (w + w').q$. If $c_{AP}$ is the operating cost of airport per passenger, the profit of airport is given by $Profit\ (\pi^{AP}) = c.f + (w + w').q - q.c_{AP}$. To absolve the model from unnecessary variables, as like Barbot (2004), we also supposed that for non-aeronautical revenues, each passenger buys one unit of goods and services at the airport, and that their price is equal to \$1. Then if $q$ is the total number of passengers, the profit equation is modified as

$$Profit\ (\pi^{AP}) = c.f + (w+1).q - q.c_{AP} \qquad (3)$$

The airport is assumed to be a utility maximiser, where utility is the sum of profit and CSR activities. CSR is the amount of welfare that the airport generates, and is a fraction $\mu$ of consumer surplus that it is willing to spend for CSR activity (Fu & Zhang, 2010). Consumer surplus (CS) in turn, is the difference between the worth/value of the service consumers receive and the price that they pay. Therefore, CS is given by $\int_{p_{min}}^{p_{max}} q\ dp,$ where $q$ is the demand function. CSR derived through consumer surpluses is given by $\mu.\int_{p_{min}}^{p_{max}}(q)dp$.

$$\mu \int_{p_{min}}^{p_{max}} q\ dp = \mu \int_{p_{min}}^{p_{max}} \left(\alpha - \beta p - \frac{\gamma}{f} + \xi\theta\right).dp = \frac{\mu q^2}{2\beta} \qquad (4)$$

The CS fraction $\mu$ allows us to define the objective function of a utility-maximising airport as a weighted sum of profit and CSR. $\mu=0$ signifies that the airport is a pure profit maximiser, whereas $\mu=1$ indicates that the firm is the perfect CSR maximiser. Airport's weighted objective function is given by

$$U^{AP} = (1-\mu)\pi^{AP} + \mu\frac{q^2}{2\beta} \qquad (5)$$

Thereby, we capture all three elements (economic, environmental, and social) with the airline as a profit maximiser with environmental responsibilities, and the airport as a utility maximiser with social responsibility. In section 4, we develop models for the centralised case and various other contract structures.

### 4. Centralised Case and Various Contract Structures

#### 4.1 Centralised Case

Under the centralised airport-airline structure, there is a single decision-maker that decides the optimal conveyance fees, ticket fare, and the greening level. In 'coordination' literature, the centralised case provides benchmark solution. Examples from real life include the Koh Samui Airport, Thailand, which effectively is owned and operated by Bangkok Airways. Both the airport and the airline cooperate among each other to implement joint decisions. Another example includes the terminal 5 of Kuwait International Airport, which is exclusively operated by Jazeera Airways. We model centralised structure where the objective of the central planner is to maximise the social welfare (SW), given by the sum of airline and airport objectives ($Eq.2 + Eq.5$), and find a globally optimal solution. The objective function is given as:

$$SW_{CENT} = (1-\mu)(pq - c_{AP}q - c_{AL}q + w'q - I\theta^2) + \mu q^2/2\beta \qquad (6)$$

As discussed, the optimal solutions of a centralised case act as benchmark solution for contract, which is discussed further. Importantly, the performance of contracts are evaluated using 2 parameters, i.e. social welfare efficiency, with a ratio of $SW/SW_{CENT}$ and greening efficiency, with $\theta/\theta_{CENT}$.

**4.2 Decentralised channel (D-CENT)**

We now analyse a decentralised structure by considering economic, environmental, and social dimensions. The structure represents the case wherein both the airport and the airline act as different entities with no coordination whatsoever. It is best represented by the example of Kolkata International Airport, India. Herein, in addition to the landing fees charged on a per-flight basis, the airport charges a conveyance fee of $17 for each international passenger, and $7 for each domestic passenger. The airport does provide essential services to airlines to facilitate operation, and charge a fee for its services. It announces landing fees and conveyance fees (CF) beforehand, and thus undertakes the Stackelberg leader's role. Based on this information, in the second stage, the airline as the Stackelberg follower, chooses the level of greening ($\theta$) and ticket fare ($p$). Interestingly, the notion of the airport as leader, and airline as follower has also been used by Xiao et al. (2016), while discussing an airport's capacity in airport-airline vertical coordination. Furthermore, Zhang et al. (2010) and Fu & Zhang (2010) also under the very same context, analysed the effects of airport concession revenue sharing on an airline's social welfare. The optimal solutions are thereby derived using backward induction. Herein, the airline's objective function is to maximise its profit $\pi^{AL} = pq - wq - cf - I\theta^2 - c_{AL}q$. The first and second-order conditions are given by: $\frac{d\pi^{AL}}{dp} = 0$; $\frac{d\pi^{AL}}{d\theta} = 0$; $\frac{d^2\pi^{AL}}{dp^2} < 0$; $\frac{d^2\pi^{AL}}{d\theta^2} < 0$. Taking the above-obtained value as input, the airport maximises its utility $U^{AP} = (1-\mu)\pi^{AP} + \mu q^2/(2\beta)$, to obtain the optimal CF.

**4.3 Cost-sharing contract (CSC)**

Greening investment requires considerable financial support in the initial days, with long gestation period; it could be in the form of R&D investment, and/or procurement of a new and efficient fleet. Notably, a single player with such exposure, does face the risk of capital loss. Consider the case of the Indian Low-cost carrier, Indigo. The airline updated its fleet with more fuel-efficient aircrafts, and placed the largest one-time order worth $30 billion to procure a fleet of A320neo aircrafts. However, soon after receiving the fleet, it faced engine failure due to technical glitches and was forced to ground its new aircraft. The company witnessed a decline in capacity growth up to 25% in 2019, and an estimated 10%

decline for next year (Kotoky & Philip, 2019). Thus, investments in greening has its inherent financial risks, like fault in procurement (or modification) of green aircraft or failure in R&D. High uncertainty and risks associated with new investments, do require firms to pool in their resources, and enter into joint-development contracts (Bharskaran & Krishnana, 2009).

Under such a scenario, we would like to propose a CSC solution, so that the airline shares a fraction of its greening investment cost with the airport in order to reduce the inherent risks. Let us presume that the airport also offers to share the investment cost to encourage the airline to undertake greening initiatives. Thus, it shares a fraction of the greening cost ($0 < \phi < 1$), while the airline pays the rest ($1 - \phi$). The airline's objective is given by $\pi^{AL} = pq - wq - cf - (1-\phi)I\theta^2 - c_{AL}q$, while the airport's objective is to maximise utility, given by $U^{AP} = (1-\mu)\pi^{AP} + \mu q^2/(2\beta)$, where $\pi^{AP} = cf + (w+1)q - c_{AP}q - \phi I\theta^2$. The cost-sharing fraction is obtained by solving $\frac{dU^{AP}}{d\phi} = 0$.

### 4.4 Revenue-sharing contract (RSC)

In RSC between two players, the downstream player agrees to share its revenue with the upstream player, in return for a lower price for the service/product procured. We model RSC such that an airline shares $\psi$ fraction of its revenue with the airport in return of lower CF charges by the airport. The remaining $1 - \psi$ of the revenue is kept by the airline. This decrease in airport charges, helps the airline to divert its available funds to make greening investments. Under this contract, both airport and airline share uncertainties and risks of green investment. Herein notably, the airlines maximises its profit, as represented by $\pi^{AL} = \psi pq - wq - cf - I\theta^2 - c_{AL}q$; while the airport maximises 'utility', as given by $U^{AP} = (1-\mu)\pi^{AP} + \mu\frac{q^2}{2\beta}$, where $\pi^{AP} = cf + (w+1)q - c_{AP}q + (1-\psi)pq$.

### 4.5 Linear two-part tariff (LTT)

We now propose an LTT contract between an airport and an airline with an economic, environmental, and social dimensions. Consider the example of the Dallas International Airport, US, where the revenue that exceeds the airport's operating cost is shared among signatory airlines. The airport keeps only a fixed revenue to cover its costs. We examine a similar case; however, in our model, the airline retains its reserved profit, and transfers the rest to the airport as a lump-sum tariff. Since the airline undertakes greening investment by entering into LTT agreement with the airport, it reduces the risk of its profit decline, which may occur due to uncertainties in return from green investments. Hence, the LTT contract helps the airline to earn a constant profit, while undertaking the financial risk of greening.

To find the optimal ticket fare and greening level, we solve the airline's objective function, given by $\pi^{AL} = pq - wq - cf - I\theta^2 - c_{AL}q - L$, while to find optimal CF and lump-sum tariff to be paid, we solve the airport's objective, given by $U^{AP} = (1-\mu)\pi^{AP} + \mu q^2/(2\beta)$, where $\pi^{AP} = cf + (w+1)q - c_{AP}q + L$.

## 5. Analytical Results

In this section, we provide analytical insights to illustrate the findings of our paper. We solve the cases discussed above to obtain the optimal equilibrium values (Table 3). we formulate a two-stage Stackelberg game involving the airport, and the airline. Airport is the leader and decide the conveyance fee, after which the airline decides the ticket fare and greening level. The model is solved using backward induction. The detailed solution and steps are provided in online appendix.

We find that under the centralised case, the condition for joint concavity of $SW_{CENT}(p,\theta)$ are: (i) $I > \frac{\xi^2(1-\mu)}{2\beta(2-3\mu)}$ (ii) $\mu \in \left(0, \frac{2}{3}\right)$. Condition (i) provides a lower limit for the greening cost $(I)$, and signifies that the airline would have to invest a certain minimum amount to achieve greening. It suggests firms to take higher greening actions if passengers are concerned for the environment. Hence, in countries where passengers make conscious decisions in favor of environmental-friendly services, the airline would tend to invest more in green technologies. The second term $\mu \in (0, \frac{2}{3})$ provides the range of CS coefficient for which the condition of concavity holds.

From the equilibrium result of the centralised case, we observe that, $\frac{(\partial SW^*)}{\partial \mu} > 0$, if $\frac{2\beta I - \xi^2}{6\beta I - \xi^2} < \mu$ and the channel profit $\frac{\partial \pi^*}{\partial \mu} < 0$, if $\frac{4\beta I - \xi^2}{6\beta I - \xi^2} > \mu$. The solution implies that when airline increases its focus on CSR, the channel profit decreases but social welfare increases. We also find that the rate of increase in total social welfare with μ is higher than the rate of decrease in profit, i.e. $\left|\frac{\partial SW^*}{\partial \mu}\right| > \left|\frac{\partial \pi^*}{\partial \mu}\right|$, under the condition $0 < \mu < 0.2$. The results are calculated by assuming passenger greening sensitivity $(\xi)$ and schedule delay $(\gamma)$ to be negligible to simplifying the analytical result.

## Table 3: Optimal Results

| | CENT | D-CENT | CSC | RSC | LTT |
|---|---|---|---|---|---|
| $\theta$ | $\dfrac{\xi(1-\mu)\Delta_1}{f(\Delta_2+2\beta I\mu)}$ | $\dfrac{\xi(1-\mu)\Delta_1}{2f(\Delta_2+3\beta I\mu)}$ | $\dfrac{6\xi(1-\mu)\Delta_1}{f\Delta_{10}}$ | $\dfrac{\xi(1-\mu)\Delta_1}{f(\Delta_2+2\beta I\mu)}$ | $\dfrac{\xi(1-\mu)\Delta_1}{f(\Delta_2+2\beta I\mu)}$ |
| $q$ | $\dfrac{2\beta I(1-\mu)\Delta_1}{f(\Delta_2+2\beta I\mu)}$ | $\dfrac{\beta I(1-\mu)\Delta_1}{f(\Delta_2+3\beta I\mu)}$ | $\dfrac{8\beta I(1-\mu)\Delta_1}{f\Delta_{10}}$ | $\dfrac{2\beta I(1-\mu)\Delta_1}{f(\Delta_2+2\beta I\mu)}$ | $\dfrac{2\beta I(1-\mu)\Delta_1}{f(\Delta_2+2\beta I\mu)}$ |
| $w$ | | $\dfrac{\Delta_4}{2\beta f}-\dfrac{I\mu\Delta_1}{2f(\Delta_2+3\beta I\mu)}$ | $\dfrac{\Delta_6}{f\beta^2 I\Delta_{10}}$ | $\dfrac{2I\mu\psi\gamma-f\Delta_8}{f(\Delta_2+2\beta I\mu)}$ | $\dfrac{\Delta_9(1-\mu)-2I\mu\Delta_{11}}{f(\Delta_2+2\beta I\mu)}$ |
| $p$ | $\dfrac{\Delta_3}{(\Delta_2+2\beta I\mu)}$ | $\dfrac{f\Delta_5-(\Delta_2+2\beta I\mu)}{2\beta(\Delta_2+3\beta I\mu)}$ | $\dfrac{f\Delta_7-\gamma(3\Delta_2+4\beta I(3-2\mu))}{\beta f\Delta_{10}}$ | $\dfrac{\Delta_3}{(\Delta_2+2\beta I\mu)}$ | $\dfrac{\Delta_3}{(\Delta_2+2\beta I\mu)}$ |
| | - | - | $\phi=\dfrac{1}{3}$ | - | $L=\dfrac{\Delta_{13}}{f^2(\Delta_2+2\beta I\mu)^2}$ |
| $\Pi^{AL}$ | | $\dfrac{I(-\xi^2+4\beta I)(1-\mu)^2\Delta_1^2}{4f^2(\Delta_2+3\beta I\mu)^2}-cf$ | $\dfrac{8I\Delta_1(f\Delta_7-3\Delta_2-4\beta I(3-2\mu))(1-\mu)}{f^2\Delta_{10}^2}-\dfrac{8\beta c_{AL}I\Delta_1(1-\mu)}{f\Delta_{10}}-\dfrac{8\Delta_1\Delta_6(1-\mu)}{\beta f^2\Delta_{10}^2}-\dfrac{24I\Delta_1^2\xi^2(1-\mu)^2}{f^2\Delta_{10}^2}-cf$ | $\dfrac{2\beta I\Delta_1\Delta_4\psi(1-\mu)}{f(\Delta_2+2\beta I\mu)^2}-\dfrac{2\beta c_{AL}I\Delta_1(1-\mu)}{f(\Delta_2+2\beta I\mu)}-\dfrac{I\Delta_1(1-\mu)\left(\begin{array}{c}\Delta_1\xi^2(1-\mu)\\+2\beta(f\Delta_8-2I\mu\psi\gamma)\end{array}\right)}{f^2(\Delta_2+2\beta I\mu)^2}-cf$ | $\bar{\pi}^{AL}$ |
| $\Pi^{AP}$ | - | $\dfrac{cf}{+\dfrac{I\Delta_1(\Delta_2+3\beta I\mu)(1-\mu)(\Delta_5+2\beta f)}{-\beta I^2\mu\Delta_1^2(1-\mu)}}{2f^2(\Delta_2+3\beta I\mu)^2}-\dfrac{\beta c_{AP}I\Delta_1(1-\mu)}{f(\Delta_2+3\beta I\mu)}$ | $\dfrac{8\Delta_1(fI(9\Delta_2+4\beta I(8\mu-1))\beta^2+\Delta_9)(1-\mu)}{\beta f^2\Delta_{10}^2}-\dfrac{12I\Delta_1^2\xi^2(1-\mu)^2}{f^2\Delta_{10}^2}-\dfrac{8\beta c_{AP}I\Delta_1(1-\mu)}{f\Delta_{10}}+cf$ | $\dfrac{2\beta I\Delta_1(1-\mu)(f(\Delta_2+2\beta I\mu)-f\Delta_8+2I\mu\psi\gamma)}{f^2(\Delta_2+2\beta I\mu)^2}-\dfrac{2\beta c_{AP}I\Delta_1(1-\mu)}{f(\Delta_2+2\beta I\mu)}+cf+\dfrac{2\beta I\Delta_1\Delta_4(1-\mu)(1-\psi)}{f(\Delta_2+2\beta I\mu)^2}$ | $L+cf+\dfrac{2\beta I\Delta_1\left(\dfrac{\Delta_9(1-\mu)-2I\mu\Delta_{11}}{f(\Delta_2+2\beta I\mu)}+1\right)(1-\mu)}{f(\Delta_2+2\beta I\mu)}-\dfrac{2\beta c_{AP}I\Delta_1(1-\mu)}{f(\Delta_2+2\beta I\mu)}$ |
| $U^{AP}$ | - | $\dfrac{I\Delta_1(1-\mu)^2(\Delta_5+2\beta f-2\beta c_{AP}f)}{2f^2(\Delta_2+3\beta I\mu)}+cf(1-\mu)$ | $\dfrac{4\Delta_1(1-\mu)^2\Delta_{12}}{\beta f^2\Delta_{10}^2}+cf(1-\mu)+\dfrac{8\beta I\Delta_1(1-c_{AP})(1-\mu)^2}{f\Delta_{10}}$ | $\dfrac{2\beta I^2\mu\Delta_1(\Delta_1+2\psi\gamma)(1-\mu)^2}{f^2(\Delta_2+2\beta I\mu)^2}-\dfrac{2\beta I\Delta_1(1-\mu)^2\left(\begin{array}{c}\Delta_8-\Delta_4-\Delta_2+2\beta I\mu\\+c_{AP}(\Delta_2+2\beta I\mu)+\Delta_4\psi\end{array}\right)}{f(\Delta_2+2\beta I\mu)^2}+cf(1-\mu)$ | $\dfrac{2\beta I\Delta_1(1-\mu)^2\left(\begin{array}{c}(1-\mu)\Delta_9+I\mu\Delta_1\\-2I\mu\Delta_{11}\end{array}\right)}{f^2(\Delta_2-2\beta I\mu)^2}-\dfrac{2\beta I\Delta_1(c_{AP}-1)(1-\mu)^2}{f(\Delta_2-2\beta I\mu)}+(1-\mu)(L+cf)$ |
| $W$ | $\dfrac{I(1-\mu)^2(\Delta_1)^2}{f^2(\Delta_2+2\beta I\mu)}$ | $\dfrac{I(1-\mu)^2(3\Delta_2+10\beta I\mu)(\Delta_1)^2}{4f^2(\Delta_2+3\beta I\mu)^2}$ | $\dfrac{4I(1-\mu)^2(15\Delta_2+4\beta I(2-3\mu))(\Delta_1)^2}{f^2(\Delta_{10})^2}$ | $\dfrac{I(1-\mu)^2(\Delta_1)^2}{f^2(\Delta_2+2\beta I\mu)}$ | $\dfrac{I(1-\mu)^2(\Delta_1)^2}{f^2(\Delta_2+2\beta I\mu)}$ |

The values of $\Delta_i$ (for i = 1 to 13) is provided in the appendix A.

On comparing the social welfare (SW) (sum of airline and airport objectives) across contracts, we find that;

*Proposition 1: In the case of airport-airline coordination with the economic, environmental, and social considerations, the social welfare of the channel follows the order $SW_{LTT} = SW_{RSC} = SW_{CENT} > SW_{D-CENT} > SW_{CSC}$ under the condition $\frac{(3\Delta_2 + 10\beta I\mu)(9\Delta_2 + 4\beta I(8\mu-1))^2}{4(15\Delta_2 - 4\beta I(3\mu-2))(2\Delta_2 + 6\beta I\mu)^2} > 1$*

From the above proposition, we conclude that SW of the channel under RSC and LTT contracts is equal to the centralised case and is always higher than D-CENT and CSC. RSC and LTT contracts coordinate with the centralised structure, with social welfare efficiency[4] (SWE) and greening efficiency[5] (GE) of 1. For D-CENT and CSC contract, SWE and GE < 1 since $SW_{CENT} > \pi^{AL}_{D-CENT} + U^{AP}_{D-CENT} > \pi^{AL}_{CSC} + U^{AP}_{CSC}$, double marginalisation persists in D-CENT and CSC contract, and the channel lacks coordination.

*Proposition 2: In the case of airport-airline coordination with the economic, environmental, and social considerations, the optimal value of conveyance fees follows the following order: $w_{CSC} > w_{D-CENT} > w_{LTT} > w_{RSC}$ under the condition, $min\left[\frac{\Delta_9(1-\mu)+\Delta_8 f}{2I\mu(\Delta_{11}+\gamma\psi)}, \frac{\beta^2 I\mu\Delta_1\Delta_{11}}{(\Delta_2+3\beta I\mu)(f\Delta_4\Delta_{11}-2\Delta_6)}, \frac{\{\Delta_2\Delta_4 + \beta I\mu(3\Delta_4-\Delta_1)\}(\Delta_2+2\beta I\mu)}{2\beta(\Delta_2+3\beta I\mu)\{\Delta_9(1-\mu)-2I\mu\Delta_{11}\}}\right] > 1$.*

Proposition 2 indicates that the conveyance fee (CF) is highest for non-coordinating CSC followed by D-CENT. Under CSC, since the airport shares 33% of the airline's greening investment costs, it charges higher CF to cover its expenses. The cost incurred by the airport in CSC is thereby higher than D-CENT, leading to higher CF thereof. For coordinating contracts, the airline either shares a fraction of its revenue or pays a tariff to the airport in return of a reduced CF. Under RSC, the airport charges a minimum CF, followed by LTT, $w_{LTT} > w_{RSC}$. On comparing the optimal CF in D-CENT and CSC (Table 3), we find that CF is an increasing function of $c_{AP}$ and is always higher than $c_{AP}$ under the condition $\xi^2 > 3\beta I$ for D-CENT and $\mu < \frac{32\beta I - 9\xi^2}{30\beta I - 9\xi^2}$ for CSC.

*Proposition 3: For the airport-airline coordination with the economic, environmental, and social considerations, the optimal value of greening level follows the following order:*

---

[4] Social welfare efficiency (SWE) is defined as the ratio of the sum of profits of the airline and the utility of the airport divided by the social welfare of the centralized case; $\frac{U^{AP}_i + \pi^{AL}_i}{SW_{CENT}}$

[5] Greening efficiency (GE) is defined as the ratio of the greening level in any case divided by the greening level in the centralized case; $\frac{\theta_i}{\theta_{CENT}}$

$\theta_{CENT} = \theta_{LTT} = \theta_{RSC} > \theta_{CSC} > \theta_{D-CENT}$ *under the condition of* $4\beta I(2 + 5\mu) > 3(1 - \mu)\xi^2 > 2\beta I(4 - 5\mu)$.

Proposition 3 suggests that for coordinating contracts (RSC and LTT), the optimal greening level is equal to that of a centralised case, and is higher than CSC and D-CENT. When an airport and an airline agree to coordinate with each other, i.e. the airline shares its revenue in RSC, along with a sharing a lump-sum tariff in LTT in lieu of lower upfront CF, the airline can provide service with a higher greening level (Proposition 2). Otherwise, in the decentralised case, with non-coordinating contracts, the greening level is lower. Therefore, the system can only achieve economic, environmental, and social progress with appropriately chosen coordination contracts.

*Proposition 4: For the airport-airline coordination with the economic, environmental, and social considerations, the optimal value of ticket fare follows the following order:* $p_{CSC} > p_{D-CENT} > p_{CENT} = p_{LTT} = p_{RSC}$ *under the condition,* $\min\left[\frac{\{f\Delta_5 - (\Delta_2 + 2\beta I\mu)\}(\Delta_2 + 2\beta I\mu)}{2\beta f(\Delta_2 + 3\beta I\mu)\Delta_3}, \frac{\{f\Delta_7 - \gamma(3\Delta_2 + 4\beta I(3 - 2\mu))\}(\Delta_2 + 3\beta I\mu)}{\Delta_{10}\{f\Delta_5 - (\Delta_2 + 2\beta I\mu)\}}\right] > 1$, *while the optimal passenger demand is of the order* $q_{CENT} = q_{LTT} = q_{RSC} > q_{CSC} > q_{D-CENT}$ *if condition* $16\beta I > 5\xi^2$ *holds.*

From proposition 4, we observe that the ticket fare has an order similar to CF. Interestingly, it is higher for CSC and D-CENT than the centralised channel, the airline passes on the part of increased CF to passengers by charging higher ticket fares. While in CSC, higher CF leads to higher ticket fare ($p_{CSC} > p_{D-CENT}$); however, the demand is still higher than the D-CENT case ($q_{CSC} > q_{D-CENT}$), because the airline shares its greening cost with the airport in CSC, due to which its greening level rises, thereby leading to higher demand from environmentally conscious travelers (Proposition 3). The ticket fare is equal in RSC and LTT ($p_{RSC} = p_{LTT}$), but is lower as compared to CSC and D-CENT. The higher level of greening in both RSC and LTT (Proposition 3) also leads to an increase in passenger demand as compared to CSC and D-CENT. Herein, the airport earns a higher non-aeronautical revenue from increased passenger footfall, which in turn compensates its lower revenue due to lower CF (Proposition 2). Hence, the channel achieves a higher greening level and demand in case of both RSC and LTT.

As the social welfare of RSC is equal to LTT (proposition 1), we compare the profit of airport and airline for both the contracts. Analytical analysis of both RSC and LTT provides us with the conditions for which $\pi_{LTT}^{AL} \geq \pi_{RSC}^{AL}$ and the conditions for which $\pi_{LTT}^{AP} \geq \pi_{RSC}^{AP}$, and offers insights on revenue sharing fraction (proposition 5). To have analytically tractable results, we assume that $\xi \to 0$, though the following results hold for the general case also (see numerical analysis results, Figure 2c).

*Proposition 5: In LTT and RSC contract,*

a) $\pi_{LTT}^{AL} \geq \pi_{RSC}^{AL}$; $\pi_{RSC}^{AP} \geq \pi_{LTT}^{AP}$ if $\psi = \left[\frac{\bar{\pi}^{AL}+u+c.f}{\sigma_1} | u \in (-\infty, 0) \cap \left\{\left(y - \frac{\bar{\pi}^{AL}+c.f}{\sigma_1}\right)\sigma_1 \middle| y \in (-\infty, 1)\right\}\right]$

b) $\pi_{LTT}^{AP} \geq \pi_{RSC}^{AP}$; $\pi_{RSC}^{AL} \geq \pi_{LTT}^{AL}$ if $\psi = \left[\frac{\bar{\pi}^{AL}+u+c.f}{\sigma_1} | u \in (0, \infty) \cap \left\{\left(y - \frac{\bar{\pi}^{AL}+c.f}{\sigma_1}\right)\sigma_1 \middle| y \in (-\infty, 1)\right\}\right]$

*The value of $\sigma_1$ is provided in appendix A.*

Proposition 5(a) suggests a feasible range of the revenue sharing fraction $\psi$ under which, the airline would have a higher profits in LTT as compared to RSC. However, for the airport to have a higher profit in LTT as compared to RSC, the range of $\psi$ is given by proposition 5(b), which is mutually exclusive to 5(a). We observe thereby differing preferences between the choice of RSC and LTT contracts for both the airport and the airline as when $\pi_{LTT}^{AL} \geq \pi_{RSC}^{AL}$ holds, we have $\pi_{RSC}^{AP} \geq \pi_{LTT}^{AP}$ under the same condition. We also find that RSF increases with $\bar{\pi}_{LTT}^{AL}$. Notably, higher the reservation profit for the airline in LTT, greater would be the revenue shared with the airport.

Both RSC and LTT contracts between an airline and airport help in channel coordination by reducing ticket fares, increasing the demand, greening level, and improving the airline's profit. The total social welfare of the contracts is equal to that of the centralised case. Specifically, in RSC, the airline's profit depends on RSF, while in LTT, the airline only keeps its reserved profit.

We now analyse our results for the two extreme scenarios, S1 and S2. These scenarios allow us to simplify the complex results and provide better insights into the analytical solution. In *Scenario 1 (S1)* → Passengers are green sensitive and airline undertakes greening, but the airport does not consider CSR ($\mu \to 0$; $\xi \neq 0$); while in *Scenario 2 (S2)* → Airport undertakes CSR, but passengers are not sensitive to the greening activities of the airline ($\mu \neq 0$; $\xi \to 0$).

Scenario 1 (S1): In scenario 1 ($\mu \to 0$; $\xi \neq 0$) the condition for joint concavity takes the form: $I > \frac{\xi^2}{4\beta}$, which is the lower limit for greening cost ($I$). The optimal channel profit and SW obtained are equal for the centralised contract and given as; $\pi_{CENT} = SW_{CENT} = \frac{I.\Delta_1^2}{f^2(-\xi^2 + 4\beta I)}$ as $\mu \to 0$. Notably herein, the optimal SW, greening level, and passenger demand are lower as compared to the case of the airport with CSR responsibility ($\mu > 0$).

This result signifies that CSR-oriented airports help in improving channel greening and demand. For D-CENT contract, on the other hand, we find $w_{D-CENT} = \frac{(c_{AP}-c_{AL}-1)}{2} - \frac{\gamma - \alpha f}{2\beta f}$, where $CF$ is an increasing function of the airport's operating cost. Analysing CSC, we find that CF increases when the airport gives up passenger CSR, and chooses to focus only on its profit. While in the case of RSC, the optimal CF charged is given by $w_{RSC} = (c_{AL} + c_{AP} - 1)\psi - c_{AL}$. Importantly, since CF is independent of the greening sensitivity of passengers along with the investment costs, the airline increases its profit, as greening increases the demand without an increase in CF. In LTT on the other hand, when the airport is a pure profit maximiser ($\mu \to 0$), we have $w_{LTT} = c_{AP} - 1$. Importantly, we observe that for all contracts, CF is an increasing function of the airport's operating cost. The analytical description of the findings for scenario 1 is presented in below proposition 6 and 7. To further simplify the comparison and gain better insights, we assume passenger schedule delay ($\gamma$) to be negligible for proposition 6, 7 and 8. However, the following propositions hold even for the general case, as shown in numerical analysis (section 6).

*Proposition 6: In scenario 1 of airline-airport coordination, when airport gives up CSR, and passenger schedule delay is assumed to be negligible, the impact of passenger's sensitivity towards greening has the following impact:*

a) $\frac{\partial \theta_{D-CENT}}{\partial \xi} > 0, \frac{\partial \theta_{CSC}}{\partial \xi} > 0, \frac{\partial \theta_{RSC}}{\partial \xi} > 0, \frac{\partial \theta_{LTT}}{\partial \xi} > 0$

b) $\frac{\partial p_{D-CENT}}{\partial \xi} > 0, \frac{\partial p_{CSC}}{\partial \xi} > 0, \frac{\partial p_{RSC}}{\partial \xi} > 0, \frac{\partial p_{LTT}}{\partial \xi} > 0$

c) $\frac{\partial \Pi^{AL}_{D-CENT}}{\partial \xi} > 0, \frac{\partial \Pi^{AL}_{CSC}}{\partial \xi} > 0 \text{ if } \frac{(16\beta I - 9\xi^2)}{(32\beta I - 9\xi^2)^3} > 0, \frac{\partial \Pi^{AL}_{RSC}}{\partial \xi} > 0 \text{ if } \xi^2 + 4\beta I(1 - 2\psi) < 0$

d) $\frac{\partial \Pi^{AP}_{D-CENT}}{\partial \xi} > 0, \frac{\partial \Pi^{AP}_{CSC}}{\partial \xi} > 0, \frac{\partial \Pi^{AP}_{RSC}}{\partial \xi} > 0 \text{ and } \frac{\partial \Pi^{AP}_{LTT}}{\partial \xi} > 0$

*Proposition 7: In scenario 1 of airline-airport coordination, when airport gives up CSR, and passenger schedule delay is assumed to be negligible, greening cost parameter has the following impacts:*

a) $\frac{\partial \theta_{D-CENT}}{\partial I} < 0, \frac{\partial \theta_{CSC}}{\partial I} < 0, \frac{\partial \theta_{RSC}}{\partial I} < 0, \frac{\partial \theta_{LTT}}{\partial I} < 0$

b) $\frac{\partial p_{D-CENT}}{\partial I} < 0, \frac{\partial p_{CSC}}{\partial I} < 0, \frac{\partial p_{RSC}}{\partial I} < 0, \frac{\partial p_{LTT}}{\partial I} < 0$

c) $\frac{\partial \Pi^{AL}_{D-CENT}}{\partial I} < 0, \frac{\partial \Pi^{AL}_{CSC}}{\partial I} < 0 \text{ if} \frac{(16\beta I - 9\xi^2)}{(32\beta I - 9\xi^2)^3} > 0, \frac{\partial \Pi^{AL}_{RSC}}{\partial I} < 0 \text{ if } \xi^2 + 4\beta I(1 - 2\psi) < 0$

d) $\frac{\partial \Pi^{AP}_{D-CENT}}{\partial I} < 0, \frac{\partial \Pi^{AP}_{CSC}}{\partial I} < 0, \frac{\partial \Pi^{AP}_{RSC}}{\partial I} < 0 \text{ and } \frac{\partial \Pi^{AP}_{LTT}}{\partial I} < 0$

We observe that an increase in the 'greening sensitivity' of passenger, drives an airline to undertake more greening investments to improve its greening level and capture higher demand (proposition 6a). Notably, airlines with higher greening level are also able to charge higher ticket fares (proposition 6b), increasing thereby their profits (proposition 6c). Also, higher passenger demand leads to an increase in the airport's revenue (proposition 6d). However, we note that the optimal greening level decreases with higher greening costs (proposition 7a). The higher greening cost leads to decrease in ticket fare (proposition 7b) because airlines invests less in greening due to higher cost and could not charge any premium from environmental sensitive passengers, thereby airlines profit also goes down (proposition 7c). Importantly, if the airline is willing to keep a higher greening level, its total investment cost ($I\theta^2$) increases, leading to reduced profits, while a lower greening level leads to reduced airport's profit (proposition 7d) with a dip in the passengers' footfall.

Scenario 2: In scenario 2 (S2) ($\mu \neq 0; \xi \to 0$), the optimal SW for the centralised case is given by $SW_{CENT} = \frac{\Delta_1^2(\mu-1)^2}{2\beta f^2(2-3\mu)}$; the channel profit is $\pi_{CENT} = \frac{I(4\beta I - 8\beta I\mu)(1-\mu)(\Delta_1)^2}{f^2(4\beta I - 6\beta I\mu)^2}$. We find that $SW_{\mu\neq 0;\ \xi\neq 0} > SW_{\mu\neq 0;\ \xi\to 0}$; which implies that SW is maximum when the system undertakes both greening and CSR. Further, we observe that for the centralised case, channel profit is a decreasing function of CSR weightage $\mu$ ($\partial \pi_{CENT}/\partial \mu < 0$), whereas total social welfare is an increasing function of $\mu$ ($\partial SW_{CENT}/\partial \mu > 0$). The analytical results for scenario 2 is presented in below proposition 8.

*Proposition 8: In scenario 2 of airline-airport coordination, when an airport is socially responsible, passenger schedule delay is assumed to be negligible, and passengers are not green-sensitive, the following results hold true:*

a. $\frac{\partial p_{D-CENT}}{\partial \mu} < 0, \frac{\partial p_{CSC}}{\partial \mu} < 0, \frac{\partial p_{RSC}}{\partial \mu} < 0, \frac{\partial p_{LTT}}{\partial \mu} < 0$

b. $\frac{\partial w_{D-CENT}}{\partial \mu} < 0, \frac{\partial w_{CSC}}{\partial \mu} < 0, \frac{\partial w_{RSC}}{\partial \mu} < 0, \frac{\partial w_{LTT}}{\partial \mu} < 0$

c. $\frac{\partial \Pi_{D-CENT}^{AL}}{\partial \mu} > 0$, and $\frac{\partial \Pi_{RSC}^{AL}}{\partial \mu} > 0$ if $\mu < \frac{2}{3}$

d. $\frac{\partial \Pi_{D-CENT}^{AP}}{\partial \mu} < 0$, $\frac{\partial \Pi_{RSC}^{AP}}{\partial \mu} < 0$ and $\frac{\partial \Pi_{LTT}^{AP}}{\partial \mu} < 0$ if $\mu < \frac{2}{3}$

We find that CSC performs similar to the D-CENT channel when $\xi \to 0$, as the airline has no greening investment cost, and there is no cost-sharing between airport-airline. We also note that for all contracts, CF along with the ticket price decrease with the airport's CSR efforts (proposition 8a & 8b). Importantly, the decrease in ticket price may also be due to lower CF. Furthermore, we note that an increase in CSR responsibility by the airport, does lead to higher airline profit (proposition 8c). Overall, this analysis shows that when airports become more CSR-oriented, it hurts its own profitability (proposition 8d), but on the other hand, is more beneficial both for customers (due to lower ticket price) and the airline (due to lower CF).

## 6. Numerical Analysis

To further illustrate the theoretical results, in this section, we perform a numerical analysis to gain further insights by changing certain parameter values. Notably, the values are chosen in a way to satisfy the conditions of profit and utility function concavity, along with demand functions positivity (conditions are provided in the online appendix B1). The parameter chosen are also in-line with the studies of Girvin (2010) and Brueckner & Girvin (2008) and satisfy the assumptions of model. The parameters values are: $\alpha = 100$; $\beta = 0.5$; $\gamma = 0.2$; $f = 5$; $\xi = 3$; $c = 30$; $I = 12$; $c_{AL} = 20$; $c_{AP} = 45$; $\mu = 0.18$; $\psi = 0.82$; $\theta_0 = 6.5$; $\phi = 0.33$; $\bar{\pi}^{AL} = 2500$.

Using above mentioned numerical values for D-CENT contract, we obtain $SWE = 70\%$ and $GE = 45\%$, leading to lower channel performance as compared to centralised case (Figure 1). In CSC, the optimal value of $\phi$ is found to be 0.33, with $SWE = 69\%$ and $GE = 73.8\%$. Channel efficiency is highest for LTT and RSC, as both contracts coordinate, with SWE and GE equal to that of a centralised case (Figure 1). As observed from Figure 1, SW increases with higher CS coefficient ($\mu$) and greening sensitivity of passengers ($\xi$). Further, the system generates the highest SW when an airline undertakes greening, 'doing nothing' is the worst possible strategy for the airline (Table 4).

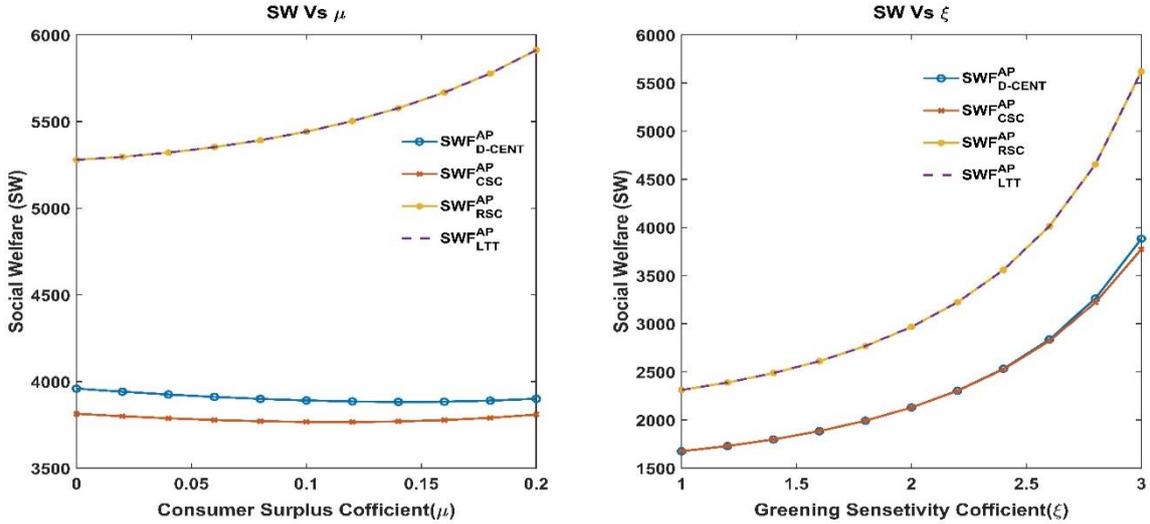

**Figure 1 (a, b): Social welfare vs. Consumer surplus coefficient and greening sensitivity coefficient**

**Table 4: Optimal Results**

| Table 3(a): Optimal solutions of the Centralised channel | | | |
|---|---|---|---|
| | BS ($\mu \neq 0$; $\xi \neq 0$) | S1 ($\mu = 0$; $\xi \neq 0$) | S2 ($\mu \neq 0$; $\xi \to 0$) |
| Total Profit | 3527.3 | 3694.8 | 2274.8 |
| Total SW | 3675.2 | 5525.2 | 2127.1 |
| Demand | 65.94 | 54.36 | 38.16 |
| Ticket fare | 166.94 | 172.73 | 163.96 |
| Greening level | 16.48 | 13.59 | 0 |

Table 4(b): Optimal solutions of the Decentralised channel

| | BS ($\mu \neq 0; \xi \neq 0$) | S1 ($\mu = 0; \xi \neq 0$) | S2 ($\mu \neq 0; \xi \to 0$) |
|---|---|---|---|
| Profit (Airline) | 960.3 | 773.9 | 496.3 |
| Utility (Airport) | 1783.7 | 1997.4 | 1124.8 |
| Demand | 29.8 | 27.18 | 17.97 |
| Ticket fare | 185.01 | 186.32 | 123.58 |
| Greening level | 7.45 | 6.7 | 0 |
| Conveyance fees | 105.4 | 111.96 | 86.76 |

Table 4(c): Optimal solutions of cost-sharing contract

| | BS ($\mu \neq 0; \xi \neq 0$) | S1 ($\mu = 0; \xi \neq 0$) | S2 ($\mu \neq 0; \xi \to 0$) |
|---|---|---|---|
| Profit (Airline) | 772.5 | 605.7 | 496.3 |
| Utility (Airport) | 1932.5 | 2147.2 | 1124.8 |
| Demand | 32.4 | 29.38 | 17.97 |
| Ticket fare | 208.03 | 207.26 | 123.58 |
| Greening level | 12.17 | 11.02 | 0 |
| Conveyance fees | 123.09 | 128.49 | 108.01 |
| Cost sharing | .33 | .33 | .33 |

Table 4(d): Optimal solutions of the revenue-sharing contract

| | BS ($\mu \neq 0; \xi \neq 0$) | S1 ($\mu = 0; \xi \neq 0$) | S2 ($\mu \neq 0; \xi \to 0$) |
|---|---|---|---|
| Profit (Airline) | 2503.1 | 1653.5 | 1834.4 |
| Utility (Airport) | 1622.6 | 2041.8 | 625.33 |
| Demand | 65.94 | 54.36 | 38.16 |
| Ticket fare | 166.94 | 172.73 | 163.96 |
| Greening level | 16.48 | 13.59 | 0 |
| Conveyance fees | 3.8 | 23.5 | 12.12 |
| Revenue-sharing fraction | 0.68 | 0.68 | 0.68 |

Table 4(e): Optimal solutions of the two-part tariff contract

| | BS ($\mu \neq 0; \xi \neq 0$) | S1 ($\mu = 0; \xi \neq 0$) | S2 ($\mu \neq 0; \xi \to 0$) |
|---|---|---|---|
| LTT Profit (Airline) | 2500 | 2500 | 2500 |
| Utility (Airport) | 1625.2 | 1194.8 | 77.05 |
| Demand | 65.94 | 54.36 | 38.16 |
| Ticket fare | 166.94 | 172.73 | 163.96 |
| Greening level | 16.48 | 13.59 | 0 |
| Conveyance fees | 15.05 | 44 | 27.24 |
| One-part tariff | 2786.6 | 1044.8 | 263.7 |

**6.1. Effects of passenger sensitivity towards greening**

Herein, we analyse the impact of greening sensitivity of passengers ($\xi$) by varying $\xi$ from 0 to 3. The change in utility and profit of the airport and airline w.r.t. $\xi$ are shown in Figure 2 (a, b, c). We observe that for a given value of parameters, the pure profit of an airport and an airline along with the utility of the airport increases over a range of $\xi$. We note that the airport earns a higher pure profit in both CSC and D-CENT contracts, as it charges higher CF (Figure 2b). For comparing the coordinating contracts, we find that the LTT contract yields a higher airport utility and profit when $\xi > 1.5$, and hence, it would be the preferred contract by the airport, wherein the customers are environment-conscious. The effect of passenger greening sensitivity on the airline's optimal profit is presented in Figure 2(c). Here, we note that for $\xi > 1.5$; $\pi_{RSC}^{AL} > \pi_{LTT}^{AL} > \pi_{D-CENT}^{AL} > \pi_{CSC}^{AL}$. Further, we observe a fall in the airline's profit when $\xi > 2.8$ for CSC, which signifies that investment in greening beyond a certain threshold, does have a negative marginal return, as the total greening cost ($I\theta^2$) increases exponentially. We conclude thereby that RSC contract performs better for the airline, as it generates higher profit, and thus would be preferred over LTT. Hence, we can affirm that channel conflicts do exist for $\xi > 1.5$, as the airport prefers LTT, as it looks to maximise both profit and utility, while the airline prefers RSC.

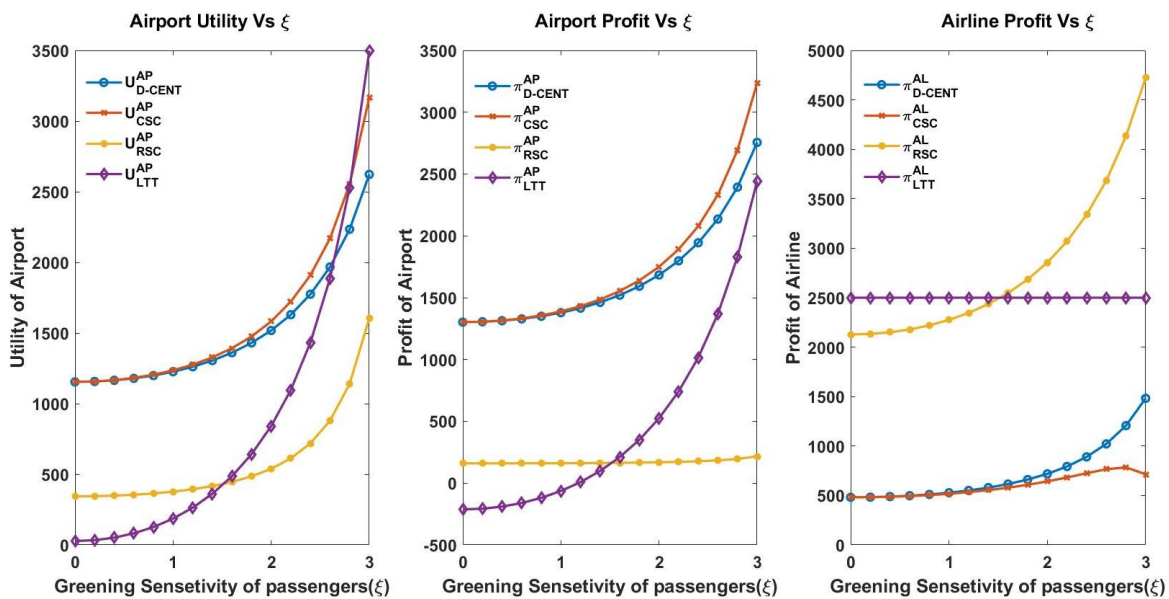

**Figure 2 (a, b, c): Airport Utility, Airport Profit, Airline Profit vs. Greening Sensitivity of Passengers**

Studying the effect of $\xi$ on CF, we find that CF decreases with increasing values of $\xi$ (Figure 3a). Since higher customer sensitivity drives airlines to improve the greening level, which requires investment, airports lend their support to the airlines by reducing CF. However, in the case of the CSC, CF increases with $\xi$, as airports already share a fraction of the increased investment costs. Due to higher CF, the greening level in CSC is lowest among all the contracts (Figure 3c). Further, we observe that in line with proposition 6, the greening sensitivity of passenger ($\xi$) does have a positive impact on decision variables $\theta$ and $p$ (Figures 3b, 3c). When $\xi$ is greater than the threshold ($\xi > 2.8$), the ticket fare in both LTT and RSC is higher than in the D-CENT (Figure 3b). When $\xi$ is equal to the threshold ($\xi = 2.8$), the ticket prices for RSC, LTT, and D-CENT are the same, but the demand is higher for coordinating contracts due to higher greening levels (Figure 3c). As passengers become environmentally conscious, the demand for green services increases with increasing greening image of the airline. Herein, the ticket fare is highest for CSC, as the airlines share 33% of greening costs with the airport, allowing thereby the latter to charge higher CF.

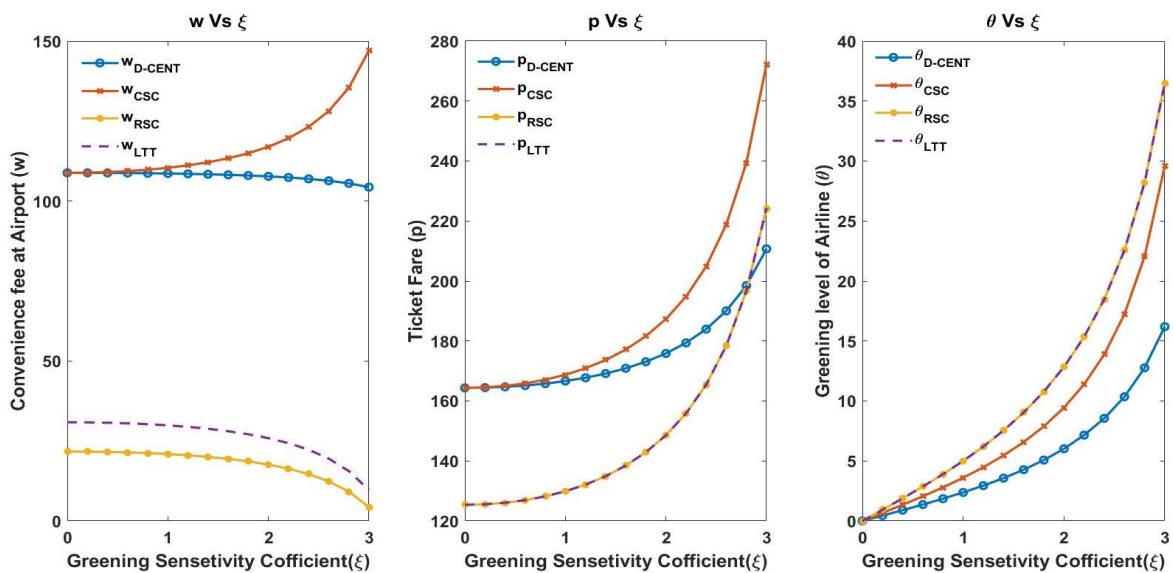

**Figure 3 (a, b, c): Conveyance fees, Ticket fare, Greening level vs Greening Sensitivity of Passengers**

### 6.2 Effects of greening cost per unit

We now explore the influence of greening cost per unit, $I$, by varying $I$ from 10 to 50 units. Figure 4 (a), (b), and (c) show the effects of $I$ on CF, the greening level and ticket fare. We observe that higher greening costs ($I$) decrease the greening level (Figure 4(b)), as the airline becomes apprehensive about investing in greening due to its higher costs. The ticket fare also decreases with higher greening costs (Figure 4c), possibly owing to the fact that as airlines invest less in greening due to higher costs, it may not be able to charge a premium from environmentally conscious passengers. The results thereby are in line with proposition 7. Further, with higher greening cost, CF increases for LTT, RSC, and D-CENT, whereas it decreases for CSC. This increase in LTT, RSC, and D-CENT may be due to the fact that, higher greening cost ($I$) leads to lower greening level, and lower demand, which in turn would decrease the airport's non-aeronautical revenue. Therefore, the airport charges a higher CF to make up for its decrease in revenue. Furthermore, in case of RSC the revenue shared by airline also decreases due to lower demand, while for LTT contract, the lump-sum tariff paid by airline decreases with $I$. Importantly, as increase in $I$ leads to reduced airline profit, the lump-sum tariff transferred to the airport decreases simultaneously. Hence, the airport increases CF in LTT, RSC and D-CENT in order to compensate for the reduced revenue (Figure 4a). For CSC, lower greening level leads to lower total cost ($I\theta^2$). Notably here, as the airport had shared a fraction of the greening cost, its own cost decreases, reducing thereby CF charged from the airline.

**Figure 4 (a, b, c): Conveyance fees, Greening level, Ticket fare vs. Greening cost parameter**

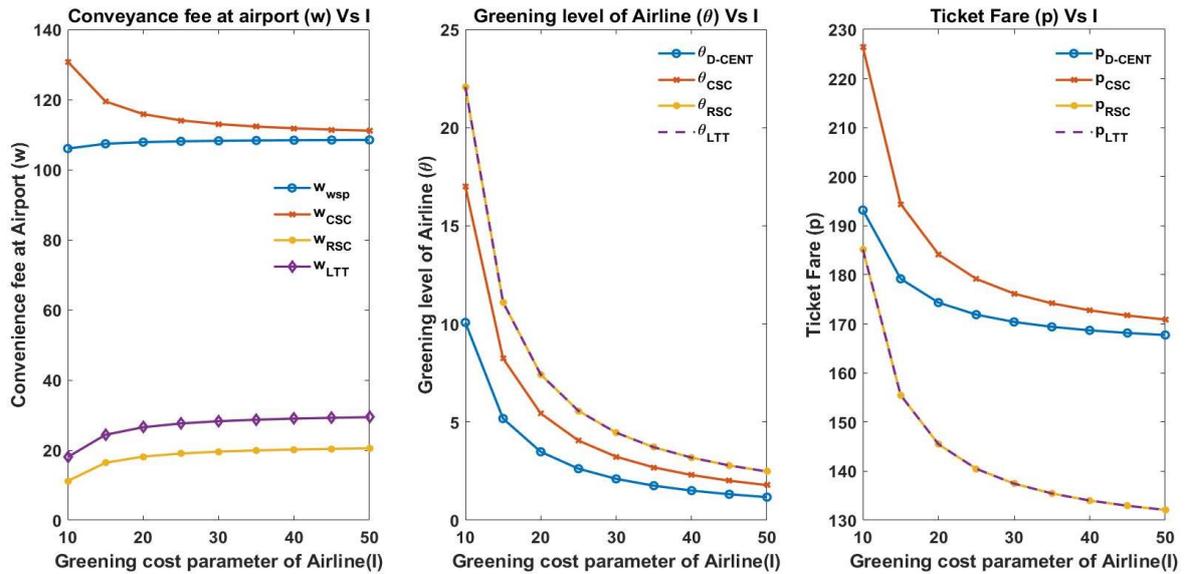

We find reduction in airport utility, airport pure profit, and airline profit with increase in per unit greening cost ($I$) (Figure 5a, 5b, 5c). Additionally, we note that in Figure 5b, when $I$ is less than a threshold, we find $\pi_{LTT}^{AP} > \pi_{RSC}^{AP}$; however, if $I$ is higher than the threshold, the profit of RS contracts is higher than LTT: $\pi_{RSC}^{AP} > \pi_{LTT}^{AP}$.

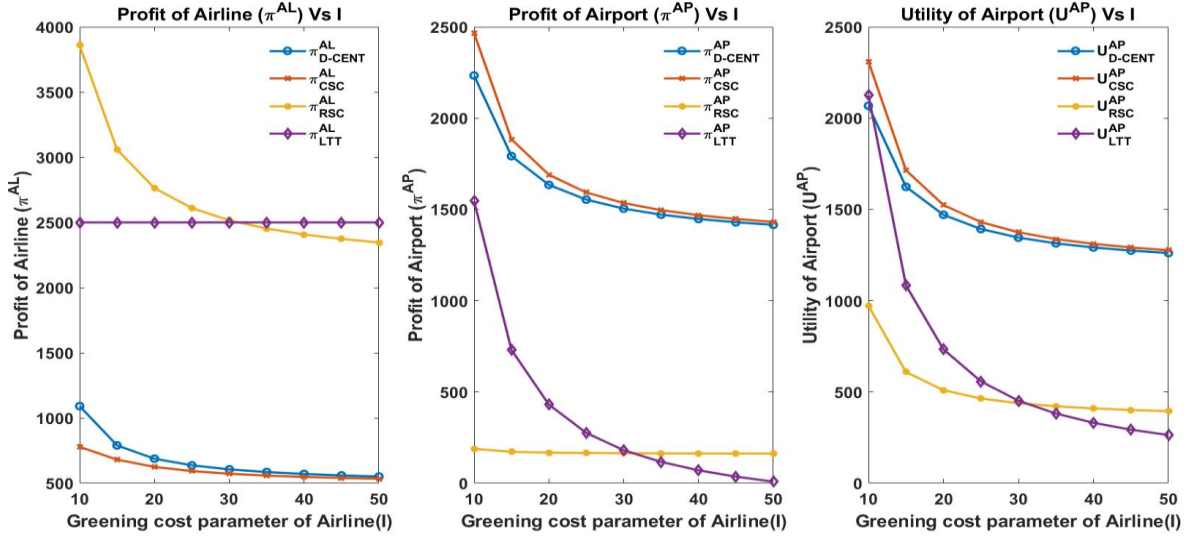

**Figure 5 (a, b, c): Profit of airline, Profit of airport, Utility of airport vs. Greening cost parameter**

### 6.3 Effect of CSR

Now, we study the impact of the airport's consumer surplus fraction μ on the performance of the contracts. When the airport puts more weight on CSR (consumer surplus), its CF charged from the airline decrease (Figure 6a), also shown in proposition 8b, $\frac{\partial w}{\partial \mu} < 0$. This reduction in CF in turn enables the airline to consider more greening, which effectively results in higher passenger demands. Moreover, due to an increase in demand at upping the greening level, the airline is able to charge a premium (Figure 6b), as environmentally conscious passengers pay higher ticket fares for greener services. This in turn allows the airline to earn a higher profit (Figure 7a). Further, we find that in LTT, airport utility ($U^{AP}$) and pure profit function ($\pi^{AP}$), are conflicting in nature; while the former increases with μ, the latter decreases (Figures 7b, 7c). For LTT, a reduction in CF by airport reduces its pure profit, but the reduction is compensated by increase in tariff that it collects from the airline. As the airline earns higher profit with increasing μ, but it keeps only the reserved profit, remainder is transferred to the airport in terms of tariff. Notably here, lower CF leads to higher passenger demand, which in turn increases airports' CSR, leading to an overall increase in $U^{AP}$.

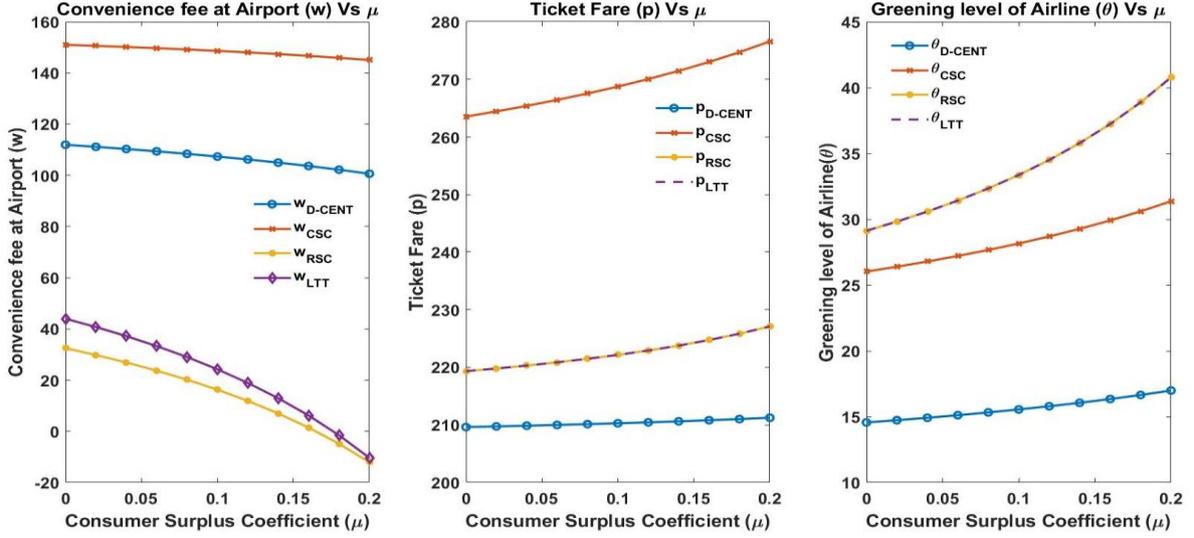

**Figure 6 (a,b,c): Conveyance fees, Ticket fare, Greening level vs. Consumer surplus coefficient**

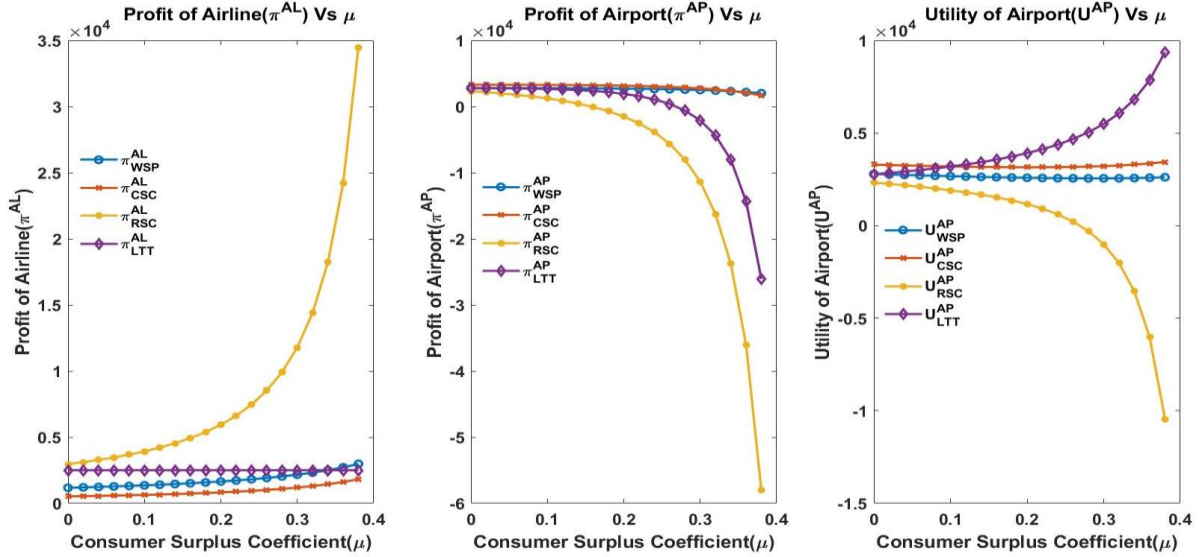

**Figure 7 (a,b,c): Profit of airline, Profit of airport, Utility of airport vs. Consumer surplus coefficient**

We also find that in the case of higher values of μ for RSC and LTT, CF can be negative, which signifies that the airport pays to the airline to invest in greening, and thereby increase its demand from environmentally-conscious passengers. However, this may not to be a practical scenario, because it destroys shareholder value with negative aeronautical profits and higher CSR. Hence, we derive conditions to have non-negative pure profit, non-negative CF, and maximum total SW. In the case of LTT, for instance, we note that the airport can earn 0 pure profit and maximise its utility for $\mu = 0.31$. For the values of $\mu > 0.31$, the airport starts incurring losses. Therefore, the airport's pure aeronautical profit is non-negative, and utility is maximum when $0 \leq \mu \leq 0.31$. In the case when the airline increases its reserved profit, the airport reduces its CSR contribution to make its pure profit non-negative. For RSC, when $\mu \leq 0.165$, the airport earns a non-negative profit.

## 7. Model Extension: Airport-airline coordination with government intervention

In this section, we extend the aforementioned model by considering the greening tax. Several governments have already started to implement stringent rules and regulations to curb emissions for the welfare of society at large (Giri et. al., 2019). In our case, the government imposes a tax on the airline to capture the environmental cost to society. The collected tax revenue is generally used to compensate the local community for environmental damages caused by the airline services. Therefore, the inclusion of the tax policy has a two-fold advantage; first, it directly targets environmental concerns by accounting for the environmental cost to society (Mankiw, 2007). Second, the emission tax may stimulate the airline to actively implement emission reduction strategies to achieve the desired level of greening. Based on this, we formulate a three-stage Stackelberg game involving the government, airport, and the airline (Figure 8).

**Figure 8: Role of government in vertical coordination**

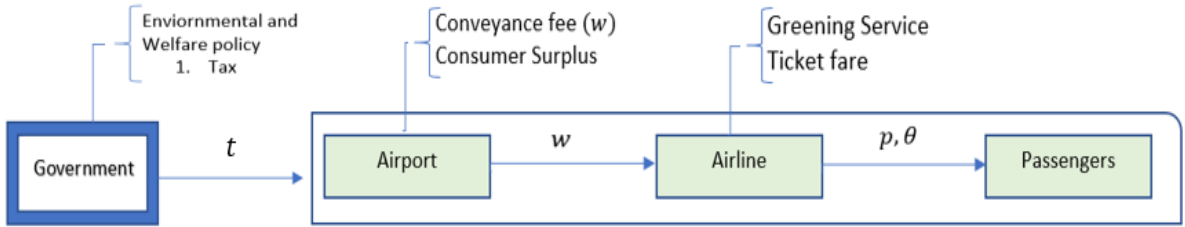

Our initial assumption of the airport as a leader and the airline the follower holds, while the government is an overall leader of the whole structure. Under the carbon tax regulation, the environment-conscious government fixes the desired level of greening ($\theta_0$) per flight operation. If the greening level of an airline is equal to $f.\theta_0$, no penalty is imposed. However, if the level of greening by the airline ($\theta$) is less than the desired level, a penalty equivalent to the difference between $f.\theta_0$ and $\theta$ is imposed. The severity of the penalty depends on the tax level coefficient $'t'$. The government tax revenue (GTR) generated from an airline for a tax level ($t \geq 0$) is given by equation:

$$GTR = t.(f.\theta_0 - \theta)^+ \tag{7}$$

The government as an overall leader decides its tax level, and based on this, the airport decides on its conveyance fee ($w$) to be charged from the airline. Finally, the airline moves last and decides on an optimal ticket fare ($p$) and greening level ($\theta$). As the airline incurs the taxation cost, the profit equation of airline is modified as:

$$Profit\ (\pi^{AL}) = (p.q) - (I.\theta^2 + c.f + w.q + q.c_{AL}) - GTR \qquad (8)$$

We analyse the four contract settings discussed in section 4, and present numerical analysis, as the obtained equilibrium results are quite large and too complex to compare analytically (refer appendix B2 for solution approach).

Figure 9: Government tax revenue vs Tax level

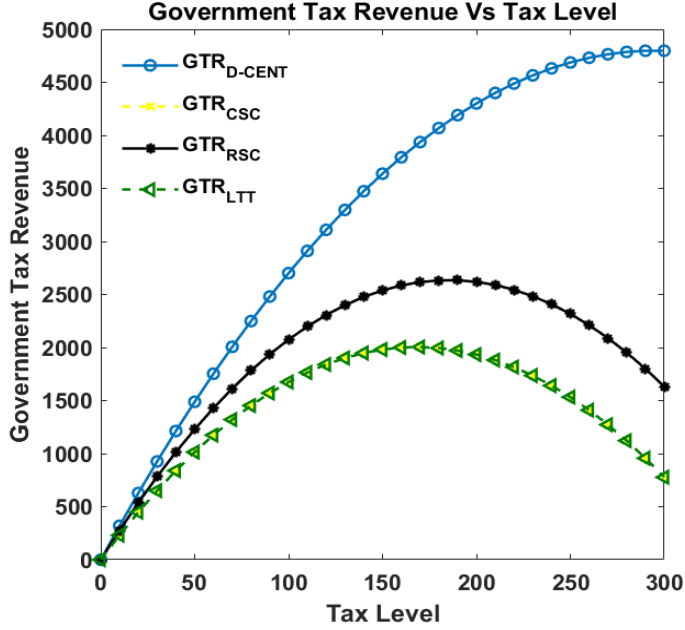

The government tax revenue (GTR) follows a concave shape (Figure 9). Herein, we observe that initially GTR increases with the tax level, reaches to a maximum level, and then begins to decline. However, the rate of increase in tax revenue and the optimal tax level is different for different contract types. If the government aims to maximise its tax revenue, there is an optimum tax level $t^*$, which would maximise its revenue. The decrease in GTR after $t^*$ maybe accounted for the fact that at higher tax levels, it may be more beneficial for the airline to invest for higher greening levels, as compared to simply paying taxes. Moreover, an increase in greening level also helps the airline by increasing its demand due to environment-conscious passengers. We can thereby achieve the same GTR at two different tax levels due to the concave down shape of the GTR function. In which case, the government would have to find a fine balance between tax revenue, desired greening level, and profitability of the airline. We discuss below the taxation for non-coordinating and coordinating contracts separately.

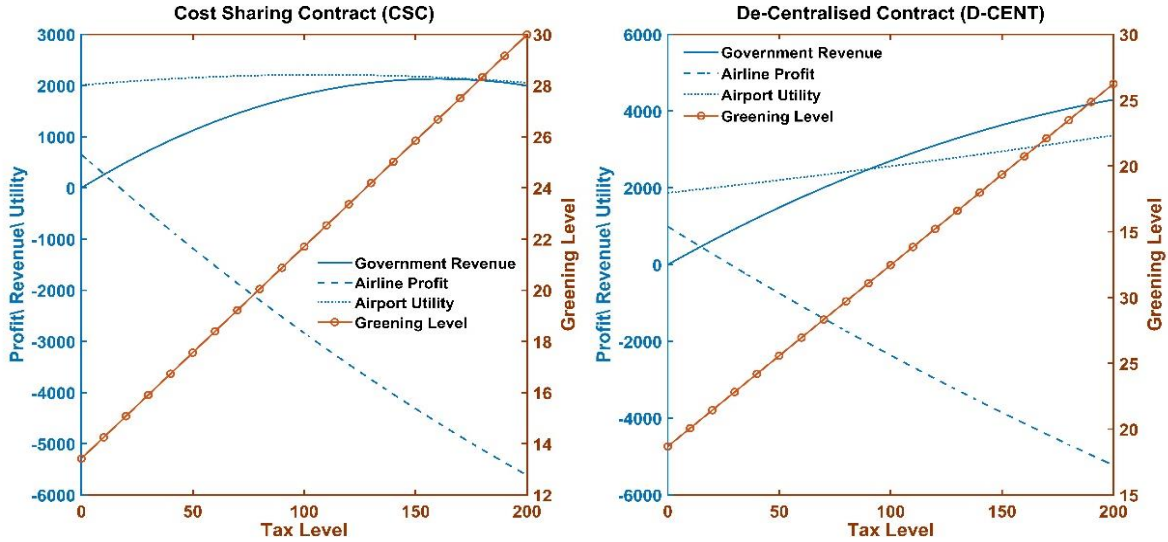

Figure 10: Non-coordinating contract vs Tax level

## 7.1. Tax in non-coordinating contracts

In this sub-section, we discuss the results obtained by imposing a tax on the airline under the non-coordinating contract cases: D-CENT and CSC. We observe a concave GTR function and an increasing greening level ($\theta$), along with an increase in the tax level (Figure 10). The ticket fare also increases with the green level (Table 5). The airport utility ($U^{AP}$) increases with tax level for D-CENT setting. However, in the case of CSC, the airport's utility curve has a downward slope at a higher tax level. This may be explained by the fact that the higher values of $\theta$ leads to higher investment costs, which needs to be shared between both the airport and the airline. The increase in greening investment cost-sharing leads in turn to a decrease in the utility of the airport. On the other hand, higher cost-sharing offsets higher revenue, cloaked by an increase in passenger demand. However, under D-CENT setting, $U^{AP}$ has an increasing slope, as the airport does not bear any greening cost, but reaps the benefits associated with an increase in demand due to environment-conscious passengers. As expected, we see a fall in the airline's profit with increasing tax level for both D-CENT and CSC contracts. The higher tax level forces the airline to undertake more greening, which further increases its investment costs. The diminishing rate of return of greening, combined with higher investment cost, along with greening tax makes the airline incur a heavy loss at higher tax levels. Hence, even though taxation helps in improving the environmental performance of the airline, and achieve the planet aspect of TBL, it destroys the economic value of the airline per se.

**Table 5. Numnerical results with greening tax**

|  | D-CENT (without tax) | D-CENT (with tax) | CSC (without tax) | CSC (with tax) | RSC (without tax) | RSC (with tax) | LTT (without tax) | LTT (with tax) |
|---|---|---|---|---|---|---|---|---|
| Tax revenue | - | 628.47 | - | 276.41 | - | 651.03 | - | 651.03 |
| Profit (Airline) | 992.07 | 82.26 | 800.25 | 485.33 | 3600.9 | 2832.8 | 2000 | 2000 |
| Utility (Airport) | 1871.2 | 2000.6 | 2026.2 | 2078.4 | 259.32 | 831.75 | 2120.1 | 1539.6 |
| Demand | 29.89 | 30.97 | 32.51 | 32.93 | 64.70 | 68.19 | 64.70 | 68.19 |
| Ticket fare | 184.97 | 189.43 | 208.04 | 208.19 | 167.56 | 173.32 | 167.56 | 173.32 |
| Greening level | 7.47 | 8.57 | 12.19 | 12.35 | 16.17 | 18.29 | 16.17 | 18.29 |
| Conveyance fees | 105.18 | 107.49 | 123.01 | 122.32 | 11.29 | 10.28 | 18.16 | 16.93 |

### 7.2. Tax in coordinating contract

Increasing tax levels leads to higher greening in coordinating contracts, i.e. RSC and LTT. However, the airport utility and airline profit have varying characteristics due to different contract properties. In RSC, the airport receives a fraction of revenue from the airline in exchange for low CF. Notably, lower CF along with higher greening levels leads to an increase in passenger demand. Subsequently, higher passenger footfall results in higher earning for the airport, from both aeronautical and non-aeronautical revenues, leading thereby to an increase in the utility of the airport. However, when the tax regime has higher greening level and ticket fare, $\theta_{RSC\_T} > \theta_{RSC}$, $p_{RSC\_T} > p_{RSC}$, we note that $\pi_{RSC}^{AL} > \pi_{RSC\_T}^{AL}$ due to tax (subscript $\_T$ represents with tax case). Also, total SW ($\pi_{RSC\_T}^{AL} + U_{RSC\_T}^{AP} + GTR_{RSC\_T}$) decreases with tax, leading to deteriorating channel efficiency.

In the case of LTT contract, the airline keeps a constant profit and transfers the rest to the airport as a lump sum tariff. Under the tax regime, it transfers lower tariffs to the airport due to lower profit and higher taxation. Hence, the utility of the airport decreases even with higher passenger footfall and greening level of the airline. In both the contracts, we observe $w > w_T$, (Table 5), which may be explained by the fact that under tax regime, the airline increases its greening, while the airport supports the airline's greening initiatives by lowering its conveyance fee. Further, the airport compensates its lower CF with an increase in revenue from non-aeronautical fees due to higher demand with greening. Meanwhile, the airline's profit decreases with a higher tax level, as it suffers a double blow from the tax and higher greening investment costs. The combined expense naturally leads the airline to become a loss-making entity.

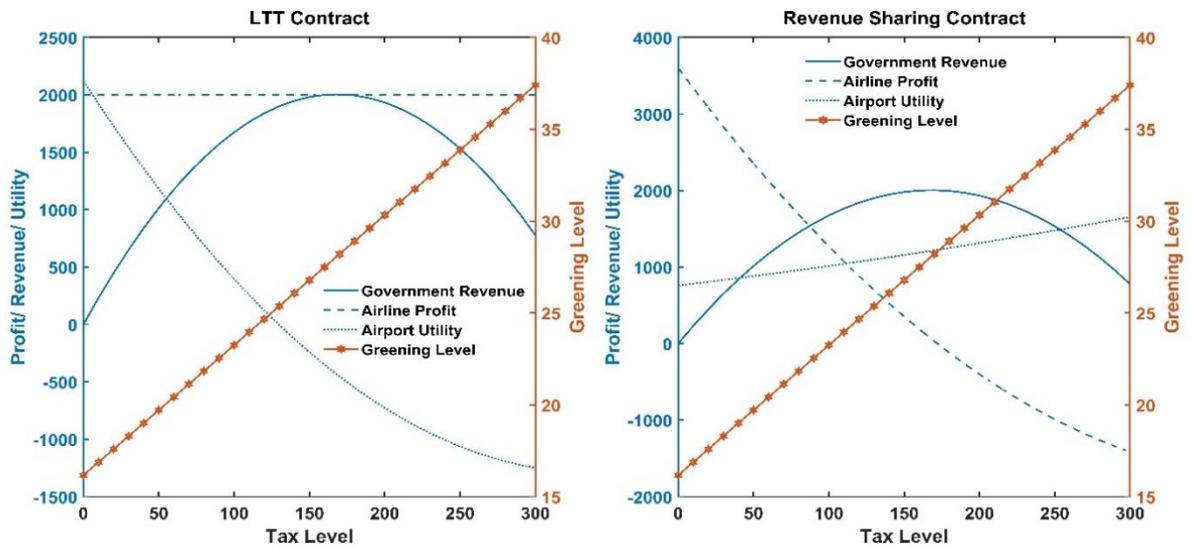

Figure 11: Coordinating contract vs Tax level

Our results have generated some interesting findings as regards government intervention. For instance, we find that the ticket fare and greening level of an airline increases with the tax level for all type of contracts. The role that tax regulation plays in terms of the greening level and welfare improvement is affected by the investment coefficient parameter of the airline. On comparing the results of coordinating contracts without tax to the decentralised case with tax, the results obtained are interesting and presented in following proposition.

*Proposition 9: We observe that $\theta_{RSC\_T} = \theta_{LTT\_T} > \theta_{RSC} = \theta_{LTT} > \theta_{CSC\_T} > \theta_{D-CENT\_T}$. The result implies that even after imposing tax in the non-coordinating contracts (D-CENT and CSC), the greening level is less than that of coordinating contracts even without tax.*

We conclude that coordination can help in achieving a higher greening level, as compared to imposing a tax in the non-coordinating cases. We also note that the airport utility increases up to a certain level of tax for almost all contracts (except in the case of LTT). The increase may be explained by the fact that the rise in greening level leads to higher passenger demands and hence higher revenue for the airport. Furthermore, the leader (airport) is able to respond to the follower's decisions of higher greening by reducing conveyance fees to support the airline to reduce its expenses in case of coordinating contracts (RSC and LTT).

Further, we show that an optimal tax level could be used as a regulatory mechanism for reducing emissions; but precautions should be taken, as it hurts the airline's profitability. The government can use the tax revenue earned for the welfare of the local community, which is affected by the negative externality of airline movements in the city. We also show that the tax revenue, profit of the airport, airline, and overall SW are all influenced by the investment coefficient parameter, tax level, and greening sensitivity of passengers. Therefore, in designing the greening tax, it is critical to consider the effect of these parameters. Moreover, the tax revenue also depends on the contract between the airport and the airline. Our results reveal that tax revenue earned under coordinating contract is lower than non-coordinating contract, as airlines undertake higher greening in RSC and LTT contract. The optimal decision on the greening level is mainly influenced by the investment cost coefficients of the airline. The optimal tax would encourage both the airport and the airline to work together to increase investments in green technologies, and to reduce emissions thereof.

### 8. Impact of greening and price competition between airlines

In this section we extend our model to include discussion of horizontal competition between two airlines. We focus on studying competition between two airlines and its effect on airport utility, airline profit, ticket fare and greening level.

### 8.1 Pricing competition between the airlines

We start our analysis by assuming pricing competition between two airlines from a single airport, providing similar services to a common market. For example, two prominent Indian domestic careers Indigo and SpiceJet operating out of their Delhi airport. We assume that both airlines make decision regarding ticket fare and greening level simultaneously. Thus, the setting deals with two heterogeneous airline and monopolistic airport (Choi, 1996).

We analyse the decentralised and revenue sharing contract case. The market size is fixed as $\alpha$, and the passenger sensitivity to airline's green service is captured by $\xi$. The pricing competition between the airlines is exhibited with different price sensitivity of passengers, represented by different substitutability coefficients $\beta_1$ and $\beta_2$. Thereby, the demand functions for the two airlines are, $q_1 = \alpha - \beta p_1 + \xi\theta_1 + \beta_1 p_2$ and $q_2 = \alpha - \beta p_2 + \xi\theta_2 + \beta_2 p_1$. $\alpha$ is the market potential, while β and $\beta_i$ is the own price and cross-price sensitivity of passengers, respectively and $0 < \beta_1 \leq \beta$ and $0 < \beta_2 \leq \beta$ holds. The demand function of airline $i$ decreases with its own ticket fare $p_i$ while increases in competitor's ticket fare $p_j$. i.e. passenger's cross price sensitivity. This demand equation helps in modelling the influences of ticket fare competition of the airlines.

For decentralised case, the airlines' profit functions are $\pi_1^{AL} = (p - w)(\alpha - \beta p_1 + \xi\theta_1 + \beta_1 p_2) - I\theta_1^2$ and $\pi_2^{AL} = (p - w)(\alpha - \beta p_2 + \xi\theta_2 + \beta_2 p_1) - I\theta_2^2$. We obtain the first order derivative with respect to $p_1, p_2, \theta_1$ and $\theta_2$. On solving the equations simultaneously, we obtain the optimal ticket fare as $p_1 = \frac{w\xi^4 - 2I\xi^2(\alpha + \beta_1 w) + (\beta w + \alpha)(8\beta I^2 + 4\beta_1 I^2) - 6\beta Iw\xi^2}{(\xi^2 - 4\beta I)^2 - 4\beta_1\beta_2 I^2}$ and $p_2 = \frac{w\xi^4 - 2I\xi^2(\alpha + \beta_2 w) + (\beta w + \alpha)(8\beta I^2 + 4\beta_1 I^2) - 6\beta Iw\xi^2}{(\xi^2 - 4\beta I)^2 - 4\beta_1\beta_2 I^2}$, while the optimal greening level are $\theta_1 = \frac{\xi((\beta - \beta_1)w\xi^2 - 4\beta I(\beta w - a) - a\xi^2 + 2a\beta_1 I + 2\beta_1 Iw(\beta + \beta_2))}{(\xi^2 - 4\beta I)^2 - 4\beta_1\beta_2 I^2}$ and $\theta_2 = \frac{\xi((\beta - \beta_2)w\xi^2 - 4\beta I(\beta w - a) - a\xi^2 + 2a\beta_2 I + 2\beta_1 Iw(\beta + \beta_1))}{(\xi^2 - 4\beta I)^2 - 4\beta_1\beta_2 I^2}$. Further, for the airport, the optimal conveyance fees charged can be found by substituting equilibrium ticket fare and green level value into profit equation and solving the first order derivative.

Analysis of the result reveals the symmetry in optimal values obtained. From the optimal values of ticket fare, we observe that $p_1 - p_2 = \frac{(2I(\beta_1 - \beta_2))}{(-\xi^2 + 4\beta I + 2\beta_2 I) + 1}$, which indicates that the ratio is dependent only on difference between cross price sensitivity of passengers $\beta_1$ and $\beta_2$. Similarly, $\theta_1 - \theta_2 = \frac{\beta(2Iw - w\xi^2 + 2aI)(\beta_1 - \beta_2)}{(a + (\beta_2 - \beta)w)\xi^2 - 4\beta^2 Iw + 4a\beta I + 2I\beta_2(a + w(\beta + \beta_1)) + 1}$; wherein equal price sensitivity of passengers leads to $\frac{\theta_1}{\theta_2} = 1$ and the ratio is independent of other parameters. Thus, with equal price sensitivity, both airlines charges similar ticket fare and under similar level of greening, while $p_1 > p_2$ for $\beta_1 > \beta_2$. On further analysis, we obtain the difference of optimal ticket fare and greening level leading to following proposition:

*Proposition 10: For D-CENT contract in pricing competition between two airlines, the difference between ticket fare and greening level of two airlines is increasing function of $\beta_1$; ($\frac{\partial(p_i - p_j)}{\partial \beta_i} > 0$; $\frac{\partial(\theta_i - \theta_j)}{\partial \theta_i} > 0$), i.e. the difference between the ticket fare and greening level increases if $I > \xi^2/4\beta + 2\beta_i$ and $I > w\xi^2/2(\alpha + \beta w)$.*

For revenue sharing case, the profit function of airlines changes to $\pi_i = r_i p_i q_i - wq_i - I\xi_i^2$, while the profit function of airport is $\pi_{AP} = (w + 1)(q_1 + q_2) - c_{AP}(q_1 + q_2) + (1 - r_1)p_1 q_1 + (1 - r_2)p_2 q_2$. The optimal values of ticket fare and greening level of airlines, with assumption of equal RCF are found to be as follows $p_i = \frac{(-r_i\xi^2 + 4\beta I + 2\beta_1 I)(-r_i w\xi^2 + 2aIr_i + 2\beta Iw)}{r_i((\xi^2 - 4\beta I)^2 - 4\beta_1\beta_2 I^2)}$; $\theta_i = \frac{I\left(4\beta(\alpha r_i - \beta w) + 2\beta_1(\alpha r_i + w(\beta + \beta_j))\right)\xi - (\alpha r_i^2 - (\beta - \beta_i)r_i w)\xi^3}{(r_i\xi^2 - 4\beta I)^2 - 4\beta_i\beta_j I^2}$.

The optimal conveyance fee is obtained by substituting the equilibrium values into objective function of airport and equating first order derivative w.r.t *w* to 0.

Analysis of the result shows that the optimal conveyance fee charged by the airport decreases with higher CSR coefficient, which is in line with our above findings (Figure 12).

We also observe increase in greening level of the airlines, both in decentralised and revenue sharing case, with increasing profit while the utility of airport decreases due to higher CSR. In case of revenue sharing contract, we find increase in conveyance fees charged by the airport when RCF is low, i.e. when airlines share lower revenue. Numerical analysis by varying passenger greening sensitivity shows decrease in conveyance fees of airport with increasing greening level. We also observe increase in profit of airlines and airport with higher greening level. The profit of airport is higher in case of RSC, as airlines share a fraction of their revenue with airport, in return of lower conveyance fees as observed in figure 13.

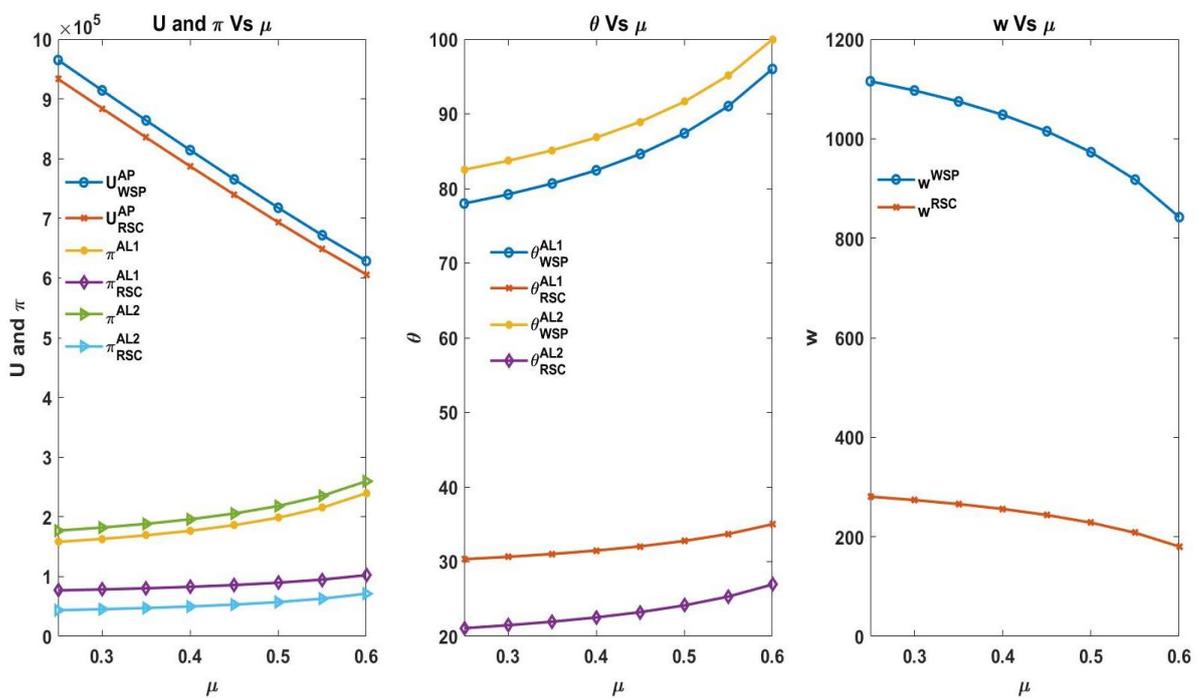

**Figure 12 (a,b,c): Profit, greening level and CF vs CSR coefficient in duoply airline market with pricing competition**

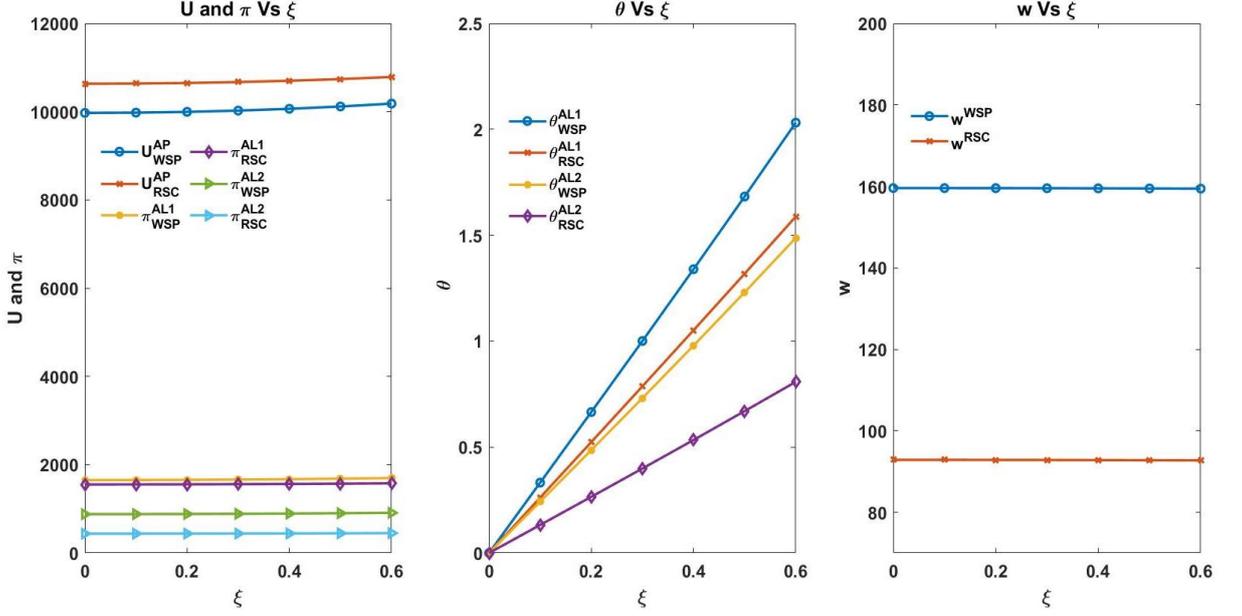

**Figure 13 (a,b,c): Profit, greening level and CF vs Greening sensitivity in duoply airline market with pricing competition**

### 8.2 Greening competition between the airlines

To further understand the impact of competition on being more environmentally conscious image, we examine the competition between two airlines undertaking greening investment under decentralised and revenue sharing case. Our previous assumptions of two competing airlines and a monopolistic airport holds. We capture the greening competition among airlines by the following demand equation $q_1 = \alpha - \beta p_1 + \xi\theta_1 - \xi_1\theta_2$ and $q_2 = \alpha - \beta p_2 + \xi\theta_2 + \xi_2\theta_1$. The market size is fixed as $\alpha$, and the passenger's price sensitivity is captured by $\beta$. The greening competition between the airlines is exhibit with different passenger's greening sensitivity, represented by different substitutability coefficients $\xi_1$ and $\xi_2$. The profit equation of airlines and airport remains the same for both the de-centralised and revenue sharing case. We solve the model using backward induction, wherein we first obtain the optimal ticket fare and greening level of airlines by first order derivative and then solve airport's objective function.

For de-centralised case, the optimal ticket fare and greening level for both the airlines are as obtained, $p_i = \frac{2I\xi(w\xi_i\beta - a(\xi+\xi_i)) + (8\beta I^2(a+w\beta) - w\xi^2(6\beta I - \xi^2 + \xi_i\xi_j))}{(\xi^2 - 4\beta I)^2 - \xi_i\xi_j\xi^2}$ and $\theta_i = \frac{\xi(w\beta-\alpha)(\xi^2 - 4I\beta + \xi_i\xi)}{(\xi^2-4\beta I)^2 - \xi_i\xi_j\xi^2}$. Analytical analysis reveals that $p_1 - p_2 = \frac{2I\xi(\xi_2-\xi_1)(\alpha-\beta w)}{16\beta^2 I^2 - 8\beta I\xi^2 + \xi^4 - \xi_1\xi_2\xi^2}$, and $\theta_1 - \theta_2 = \frac{\xi^2(\xi_2-\xi_1)(\alpha-\beta w)}{16\beta^2 I^2 - 8\beta I\xi^2 + \xi^4 - \xi_1\xi_2\xi^2}$. We observe that when $\xi_j > \xi_i$; i.e. passengers are more sensitive towards greening of airline j, the difference $\theta_1 - \theta_2$ is positive, i.e. airlines undertake higher greening and charge higher ticket fare. Further, we observe similar trend with decreasing conveyance fees and higher profit and greening level for airlines with higher CSR. It is interesting to note that utility of airport first decreases and then increases with CSR coefficient. Similar results are obtained with passenger's greening coefficient, as higher sensitivity leads to higher greening effort put by airlines. On further analysis, we obtain the difference of optimal ticket fare and greening level leading to following proposition:

*Proposition 11: For D-CENT contract in greening competition between two airlines, the difference between ticket fare and greening level of two airlines is increasing function of $\xi_1$; ($\frac{\partial(\theta_i - \theta_j)}{\partial \xi_i} > 0$), i.e. the difference increases if $4bI^2 > \xi(\xi - \xi_2)$.*

Thus, the similarity in the results obtained with duopoly airline market and single airline market under both pricing and greening competition reflect the robustness of result obtained in previous section.

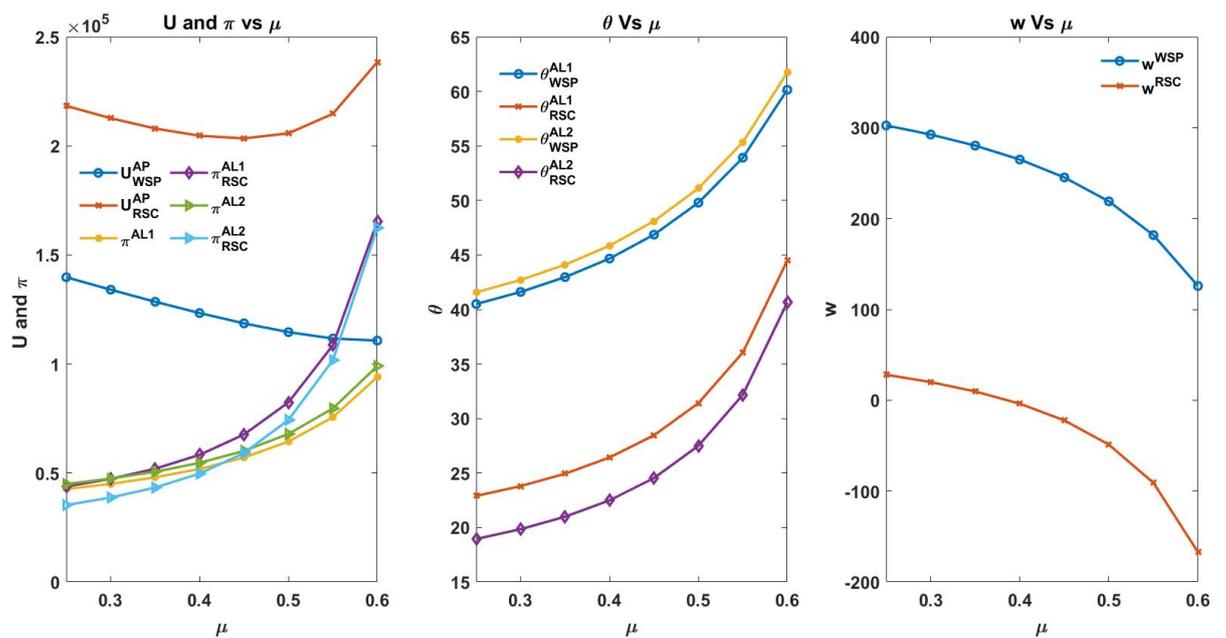

**Figure 14 (a,b,c): Profit, greening level and CF vs CSR coefficient in duoply airline market with greening competition**

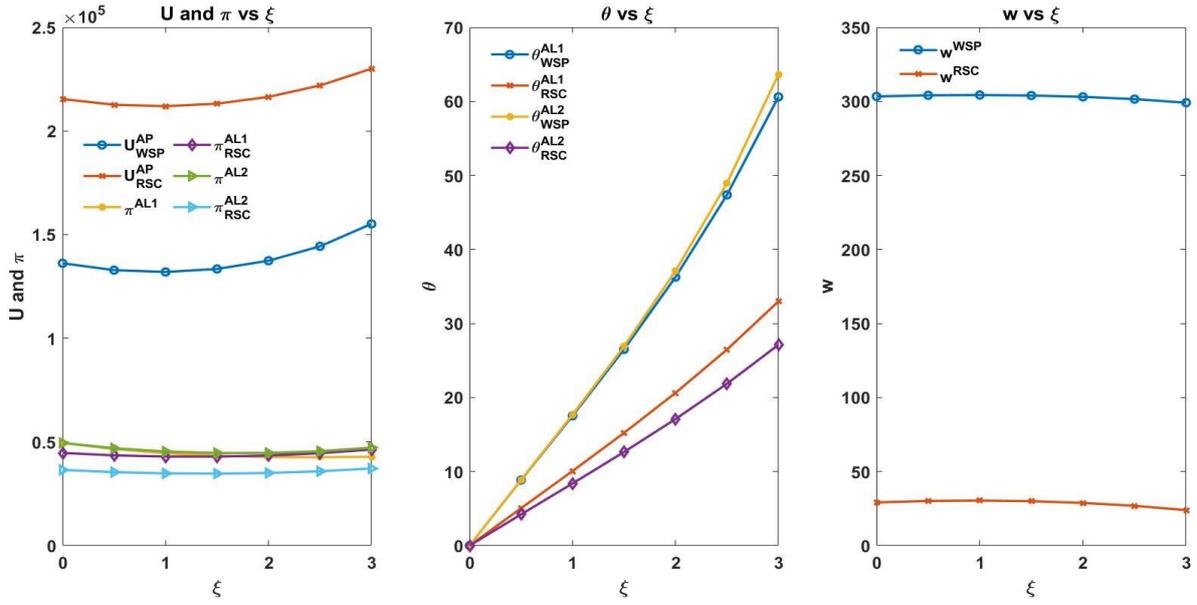

**Figure 15 (a,b,c): Profit, greening level and CF vs Greening sensitivity in duoply airline market with greening competition**

### 9. Summary and Conclusions

In this study, we considered airline-airport coordination model; with and without government intervention. The airline undertakes greening investments to provide environment-friendly service, and the airport is engaged in CSR activity. They can jointly achieve triple bottom line (TBL) growth. The government could act as an overall leader for both the airport-airline coordination model, and collect taxes for emission reduction, while catering to the welfare of the local community. Our results indicate that a linear two-part tariff and a revenue-sharing contract are best-suited to coordinate a decentralised structure by incorporating the TBL elements. These contracts provide an incentive for the airport to improve CSR, and focus on overall utility maximisation. Further, the contract provides an opportunity for the airline to achieve its environmental goals, while increasing the greening level by sharing the financial risks involved with the airport. We believe that such an analytical framework in the airline industry has not been reported before. Below we present some of the major findings of the paper along with their managerial and policy implications.

| | Table 6. Findings and Implications | |
|---|---|---|
| **Research question** | **Findings** | **Managerial / Policy implications** |
| Under what circumstances can airport-airline coordination achieve economic, environmental and social progress? | We have demonstrated the effectiveness of RSC and LTT contracts in achieving the growth. These contracts coordinate the channel like a centralised case, and improves the greening level (planet), maximises social welfare (people) and profit. We also explored CSC and D-Cent contracts and concluded that these contracts *do not* coordinate with the centralised case to provide optimal results. | The equilibrium results and propositions discussed, could help both airport and airline managers to choose the optimal values of the parameters to achieve TBL growth. Results under different scenarios that have been discussed would help managers gain better insights. |
| How can mutual support and use of contracts eliminate the problem of incoordination? | The results obtained in the study show that mutual support between the airport and an airline can improve the greening level, attract more environmentally conscious passengers, increase the profitability of the airline and utility of the airport. On comparing the results of coordinating contracts with the D-Cent case, we note that *coordination does help* both the airport and the airline to improve their utility and profit goals. The airport provides support to the airline's greening objective by lowering its charges so that the airline can utilise funds towards greening investment. In return, the airline shares a fraction of revenue (RSC) or pays a tariff (LTT) to the airport in order to compensate for its losses due to lower CF. Overall, the channel's SWE and GE increases. Resultantly, collaboration and cooperation between the airport and airline to make a cleaner environment improves, also enhancing in turn the aspect of social welfare. | When an airline undertakes greening responsibility, and the airport considers social welfare, mutual coordination in the form of LTT and RSC could help both the parties improve their objectives. A lack of coordination on the other hand, may be resolved by choosing suitable contract terms through which the airport-airline could enhance interaction with each other. Both airport & airline managers could gain valuable insights from the numerical results obtained, and thereby formulate favorable terms in line with the proposed contract solution for better coordination between the two players. For airline managers, it is viable to share the risk of greening investment with the airport to reduce the burden of greening costs by considering suitable contract terms with the airports. |
| *What is the effect of economic, environmental and social elements on factors such as airport charges, ticket fares, and demand for air travel?* | | |
| a. Impact of greening sensitivity of | The airline's investment in improving its greening (environmental) level can increase the passenger demand, the | There is a prevalent notion that greening investment might lead to a reduction in the profitability of the airline. |

**Table 6. Findings and Implications**

| Research question | Findings | Managerial / Policy implications |
|---|---|---|
| passengers and greening investment cost: | airline's own profit, and the airport's utility, which in turn enhances the social welfare. The investment cost parameter does play an important role in the airline greening level, and hence the system's performance. There is an optimal level of greening for the airline with and without government intervention, which effectively maximises its profit. In the model, with government intervention, we also have an optimal tax level, which guarantees a certain level of greening while ensuring a positive profit of the airline.<br><br>On varying passenger sensitivity, we observe higher profit for both the airport and airline with increasing sensitivity. The increasing consumer sensitivity also has an increasing impact on the ticket fare, CF, and greening level of the airline. The optimal greening level is higher for coordinating contracts and in the tax regime. However, airline's profit decreases in tax regime as it undertakes higher investment which results in a negative marginal impact. | However, the results obtained indicate that the implementation of TBL increases both the airport and the airline's profit. Therefore, managers should be motivated to put more effort into both greening and CSR. To obtain the highest level of greening, mangers should consider either LTT or RSC contract. Therefore, using the results obtained, managers can derive the optimal greening level, CF and tax level for different settings. The airline could develop an appropriate ticket price mechanism to maximise channel performance.<br><br>Results also indicate that it is viable for the airline, airport, and government to invest in increasing the consumers' awareness about the greening performance of the airline. These can be accomplished through advertisements and promotional activities. In the case of tax regime, the revenue earned by the government can be used for the welfare of the local community. The government can even help the airline in improving its greening level by providing funds to invest in environmental innovations. |
| b. Impact of CSR | We conclude that as the airport increases its CSR commitment, it offers CF below the marginal operating cost in coordinating contracts. Both contracts (RSC and LTT) achieve equal SWE, with increasing airline profit with the airport's CSR. Since the numerical results obtained are for a constant value of revenue sharing fraction, we find reduced airport profit, as the revenue shared by the airline is not enough to compensate for the loss in airport revenue due to lower CF.<br><br>Our results also show that CSR improves channel | The study reveals that for a socially concerned airport with a focus on CSR, the CF charged should be low to generate passenger demand by lower ticket prices. The airport should focus on earning profit from higher non-aeronautical revenue due to increased demand, thereby subsidising aeronautical revenue to promote greening.<br><br>However, we find a decrease in the pure profit of the airport as it focuses more on CSR. A CSR-oriented airport should decide to what extent it pays 'to be good', and beyond |

| Research question | Findings | Managerial / Policy implications |
|---|---|---|
| | coordination, as the overall SW of the system increases when the airport undertakes more CSR activities. We conclude that both environmental and social effort improves the performance of the airline-airport channel at large. | what level it should 'not to be too good'. Overall, our findings suggest that a contract can be an important source for welfare improvement based on the objective of the airport. Airports also play a role in sharing risk associated with the greening investment with the airline, and hence promote green service. The airport authorities stand to gain, as passenger footfall increase with the greening level of the airline. |

*What is the role of government intervention in achieving economic, environmental, and social progress in the aviation sector?* Finally, taxation is one policy measure that governments can use for carbon emissions reduction, especially in the case of non-coordinating airline-airport contract. To achieve the environmental and social progress, fundamentally, we need firms to work together under contract, and achieve higher CSR and greening level by investing in green technologies to reduce their emissions. However, in the case of inefficient contract types (D-CENT, CSC), the government can resort to taxation to promote the higher greening initiatives. In the case of efficient contract (RSC, LTT) with a higher level of greening, on one hand, taxation does increase the greening level, but on the other hand, it is detrimental to overall contract performance. Our findings indicate that firms' optimal decision on green technology investment is mainly influenced by their emission cost reduction strategy. Furthermore, from the perspective of policymakers, our research findings provide interesting insights into how the tax level can affect the economic and environmental performances of the entire structure. As opposed to the belief of a single tax strategy for all our findings demonstrate it is efficient to have it different for contract types. Unique characteristics of the contracts (e.g. cost-sharing or revenue sharing) should be considered in the development of the taxation scheme. Furthermore, good use of tax revenue is equally important to reduce the externality faced by the local community in the form of poor air quality and higher noise levels due to airline operations. When deciding the use of revenue, policymakers should consider the greening investment of airlines and develop incentives for airlines to invest in green technologies that can help to further reduce carbon emissions. Our findings also suggest a fundamental trade-off between economic efficiency and environmental sustainability. It is crucial for policymakers to balance this and other trade-offs considering the importance of air travel, and the immediate and the long-term economic and environmental challenges of the aviation sector at large. Finally, taxation should be used in combination with coordinating contracts. Based on our analysis,

| Table 6. Findings and Implications |||
|---|---|---|
| Research question | Findings | Managerial / Policy implications |
| | we note that it is possible to achieve a higher level of greening in coordinating contracts even without tax, as opposed to the non-coordinating contracts with tax. ||
| *How does the contract affect the duopoly airline market with greening and pricing competition to achieve better economic, environmental, and social results?* | We find that the effect of CSR coefficient and greening sensitivity on a duopoly airline market is similar to the one observed in the monopolistic case. However, in the case of price competition, the optimal ticket fare and greening level are function of passengers' cross-price sensitivity and cross greening sensitivity in greening competition. In both cases, we find a similar trend in profit, greening level, and airline ticket fare. For D-CENT contract in pricing competition between two airlines, the difference between ticket fare and greening level of two airlines is increasing function of cross price sensitivity. For greening competition, airline should undertake higher greening and charge higher ticket fares if passengers have higher cross greening elasticity. | The extended model helps to test the robustness of our result. We find similar results in the case of pricing and greening competition. From the managerial point of view, results indicate that airlines should improve their greening level as passengers become conscious, even in competitive markets. They can charge premium ticket fare for greening. Results also suggest that it is viable for the airline to increase the consumers' awareness about the greening performance. The airport should charge lower CF, as it aids airlines to focus on greening. Also, a higher CSR focus leads to a lower profit for the airport. |

## 10. Future Research Directions

Future research could explore our proposed models by relaxing some assumptions, such as competition between the airports and airlines. We have considered only one tax structure. Future studies could explore different tax structures such as cap and trade, and the combined effect of coordination. We also recommend exploring the results in stochastic demand and a multi-period setting. Importantly, in addition to the findings above, we do acknowledge the importance of factors such as trust, the risks involved, along with the issues of implementation in choosing the best contract.

**References:**

Auerbach, S., & Koch, B. (2007). Cooperative approaches to managing air traffic efficiently—The airline perspective. *Journal of Air Transport Management*, 13(1), 37–44.

Barbot, C. (2004). Economic effects of re-allocating airports slots: a vertical differentiation approach. *Journal of Air Transport Management*, 10(5), 333-343.

Barbot, C. (2009). Airport and airlines competition: Incentives for vertical collusion. *Transportation Research Part B: Methodological*, 43(10), 952–965.

Baxter, G. (2020). The Use of Aviation Biofuels as an Airport Environmental Sustainability Measure: The Case of Oslo Gardermoen Airport. *MAD-Magazine of Aviation Development*, 8(1), 6-17.

Bhaskaran, S. R., & Krishnan, V. (2009). Effort, revenue, and cost sharing mechanisms for collaborative new product development. *Management Science*, 55(7), 1152-1169.

Bitzan, J. D., & James, H. (2017). Peoples (2017). 'Introduction', The Economics of Airport Operations (Advances in Airline Economics, Volume 6) (pp. 1-14). *Emerald Publishing Limited.*

Brueckner, J. K., & Girvin, R. (2008). Airport noise regulation, airline service quality, and social welfare. *Transportation Research Part B: Methodological*, 42(1), 19-37.

Carney, M., & Mew, K. (2003). Airport governance reform: A strategic management perspective. *Journal of Air Transport Management*, 9(4), 221–232.

Caro, F., Corbett, C. J., Tan, T., & Zuidwijk, R. (2013). Double counting in supply chain carbon foot printing. *Manufacturing & Service Operations Management*, 15(4), 545-558.

Chang, Y. T., Park, H. K., Zou, B., & Kafle, N. (2016). Passenger facility charge vs. airport improvement program funds: A dynamic network DEA analysis for US airport financing. *Transportation Research Part E: Logistics and Transportation Review*, 88, 76-93.

Choo, Y. Y. (2014). Factors affecting aeronautical charges at major US airports. *Transportation Research Part A: Policy and Practice,* 62, 54-62.

Czerny, A. I., & Zhang, A. (2015). How to mix per-flight and per-passenger based airport charges. *Transportation Research Part A: Policy and Practice*, 71, 77-95.

D'Alfonso, T. (2012). Vertical relations between airports and airlines: theory and implications. PhD Dissertation.

Dray, L., Evans, A., Reynolds, T., Schäfer, A. W., Vera-Morales, M., & Bosbach, W. (2014). Airline fleet replacement funded by a carbon tax: an integrated assessment. *Transport Policy,* 34, 75-84.

Elkington, J. (1998). Accounting for the triple bottom line. *Measuring Business Excellence*, 2(3), 18–22.

Francis, G., Humphreys, I., & Ison, S. (2004). Airports' perspectives on the growth of low-cost airlines and the re-modeling of the airport–airline relationship. *Tourism Management*, 25(4), 507–514.

Fu, X., Oum, T. H., & Zhang, A. (2010). Air transport liberalisation and its impacts on airline competition and air passenger traffic. *Transportation Journal*, 24-41.

Fu, X., & Zhang, A. (2010). Effects of airport concession revenue sharing on airline competition and social welfare. *Journal of Transport Economics and Policy (JTEP)*, 44(2), 119-138.

Ghosh, D., & Shah, J. (2012). A comparative analysis of greening policies across supply chain structures. *International Journal of Production Economics*, 135(2), 568-583.

Gillen, D., & Morrison, W. (2003). Bundling, integration and the delivered price of air travel: Are low cost carriers full service competitors? *Journal of Air Transport Management*, 9(1), 15–23.

Giri, R. N., Mondal, S. K., & Maiti, M. (2019). Government intervention on a competing supply chain with two green manufacturers and a retailer. *Computers & Industrial Engineering*, 128, 104-121.

Girvin, R. (2010). Aircraft noise regulation, airline service quality, and social welfare: The monopoly case. *Journal of Transport Economics and Policy (JTEP)*, 44(1), 17-35.

IEA (2019), Are aviation biofuels ready for take-off?, IEA, Paris *https://www.iea.org/commentaries/are-aviation-biofuels-ready-for-take-off*

Kotoky, A., & Philip, S. (2019, November 29) Why IndiGo is facing problems with A320neo planes but not GoAir. *The Print*

Laroche, M., Bergeron, J., & Barbaro-Forleo, G. (2001). Targeting consumers who are willing to pay more for environmentally friendly products. *Journal of Consumer Marketing*, 18(6), 503–520.

Lin, Y., & Lu, Y. (2015). The economic impact of different carbon tax revenue recycling schemes in China: A model-based scenario analysis. *Applied Energy*, 141, 96-105.

Lu, C. (2018). When will biofuels be economically feasible for commercial flights? Considering the difference between environmental benefits and fuel purchase costs. *Journal of cleaner production*, 181, 365-373.

Mankiw, N. G. (2007). One answer to global warming: a new tax. *New York Times*, 16.

Masiol, M., & Harrison, R. M. (2014). Aircraft engine exhaust emissions and other airport-related contributions to ambient air pollution: A review. *Atmospheric Environment*, 95, 409–455.

Noëth, B. (2019, December 16). Munich Airport and Lufthansa agree to extend satellite of Terminal 2. *www.aviation24.be*

Oum, T. H., Zhang, A., & Zhang, Y. (2004). Alternative forms of economic regulation and their efficiency implications for airports. *Journal of Transport Economics and Policy*, 38(2), 217-246.

Richard, O. (2003). Flight frequency and mergers in airline markets. *International Journal of Industrial Organization*, 21(6), 907-922.

Ryder, B (2014, August 30). A new green wave. *The Economist*.

Sheu, J. B., & Li, F. (2014). Market competition and greening transportation of airlines under the emission trading scheme: a case of duopoly market. *Transportation Science*, 48(4), 684-694.

Stalnaker, T. (2019, April 25). Margins Are Tightening for US Airlines as Capacity Growth Keeps Outpacing GDP. *Forbes*.

Starkie, D. (2001). Reforming UK Airport Regulation. *Journal of Transport Economics and Policy*, 35(1), 119–135

Tsai, W. H., Lee, K. C., Liu, J. Y., Lin, H. L., Chou, Y. W., & Lin, S. J. (2012). A mixed activity-based costing decision model for green airline fleet planning under the constraints of the European Union Emissions Trading Scheme. *Energy*, 39(1), 218-226.

Walley, N., & Whitehead, B. (1994). It's not easy being green. *Reader in Business and the Environment*, 36(81), 4.

Xiao, Y., Fu, X., & Zhang, A. (2016). Airport capacity choice under airport-airline vertical arrangements. *Transportation Research Part A: Policy and Practice,* 92, 298-309.

Yaacob, M. R., & Zakaria, A. (2011). Customers' awareness, perception and future prospects of green products in Pahang, Malaysia. *The Journal of Commerce*, 3(2), 1

Yang, H., Zhang, A., & Fu, X. (2015). Determinants of airport–airline vertical arrangements: analytical results and empirical evidence. *Journal of Transport Economics and Policy*, 49(3), 438-453.

Zhang, A., & Czerny, A. I. (2012). Airports and airlines economics and policy: An interpretive review of recent research. *Economics of Transportation,* 1(1-2), 15-34.

Zhang, A., Fu, X., & Yang, H. G. (2010). Revenue sharing with multiple airlines and airports. *Transportation Research Part B: Methodological*, 44(8-9), 944-959.

Zhang, A., & Zhang, Y. (1997). Concession revenue and optimal airport pricing. *Transportation Research Part E: Logistics and Transportation Review,* 33(4), 287–296.

Zhang, A., & Zhang, Y. (2003). Airport charges and capacity expansion: effects of concessions and privatisation. *Journal of Urban Economics,* 53(1), 54-75.

Appendix A: Alias used in the paper

$\Delta_1 = \{\alpha f - \beta f(c_{AP} + c_{AL} - 1) - \gamma\}, \Delta_2 = (4\beta I + \mu\xi^2 - \xi^2 - 8\beta I\mu), \Delta_3 = \left[(1 - \mu)\{2\alpha I + (\xi^2 - 2\beta I)(1 - c_{AP} - c_{AL})\} - 2I\alpha\mu - \frac{2I\gamma}{f}(1 - 2\mu) - 2I\alpha\mu\right], \Delta_4 = \{-\gamma + \alpha f - \beta f(1 - c_{AP} + c_{AL})\}, \Delta_5 = (1 - \mu)\{(2\beta I - \xi^2)(c_{AP} + c_{AL} - 1) - \alpha\xi^2\} + 2\alpha I(3 - 4\mu), \Delta_6 = 4f\beta I(1 - \mu)(8\beta I - 3\xi^2)\{\gamma + (\beta c_{AL} + \beta c_{AP} - \alpha - \beta)f\} + (\gamma - \alpha f + \beta c_{AL}f)\{8\beta^2 I^2 \mu - 3\beta I \xi^2 (1 - \mu)\}, \Delta_7 = (1 - \mu)\{2\beta(4\beta I - 3\xi^2)(c_{AL} + c_{AP} - 1) + 3\alpha(8\beta I - \xi^2)\} - 8\beta I\mu\{\alpha + \beta(c_{AP} + c_{AL})\}, \Delta_8 = (1 - \mu)(4\beta I - \xi^2)\{c_{AL} - \psi(c_{AP} + c_{AL} - 1)\} - 2I\mu(\beta c_{AL} - \alpha\psi), \Delta_9 = (f\xi^2 - 4\beta fI - c_{AP}f\xi^2 + 4\beta c_{AP}fI), \Delta_{10} = 9\Delta_2 + 4\beta I(8\mu - 1), \Delta_{11} = \alpha f - \beta c_{AL}f - \gamma, \Delta_{12} = (2\Delta_6 - 3\beta I\Delta_1\xi^2(1 - \mu) + 8\beta^2 I^2 \mu\Delta_1),$

$\Delta_{13} = (4\alpha^2\beta f^2 I^2 \mu^2 - 8\alpha^2\beta f^2 I^2 \mu + 4\alpha^2\beta f^2 I^2 - \alpha^2 f^2 I\mu^2\xi^2 + 2\alpha^2 f^2 I\mu\xi^2 - \alpha^2 f^2 I\xi^2 - 8\alpha\beta^2 c_{AP} f^2 I^2 \mu^2 + 16\alpha\beta^2 c_{AP} f^2 I^2 \mu - 8\alpha\beta^2 c_{AP} f^2 I^2 - 8\alpha\beta^2 c_{AL} f^2 I^2 \mu^2 + 16\alpha\beta^2 c_{AL} f^2 I^2 \mu - 8\alpha\beta^2 c_{AL} f^2 I^2 + 8\alpha\beta^2 f^2 I^2 \mu^2 - 16\alpha\beta^2 f^2 I^2 \mu + 8\alpha\beta^2 f^2 I^2 + 2\alpha\beta c_{AP} f^2 I\mu^2\xi^2 - 4\alpha\beta c_{AP} f^2 I\mu\xi^2 + 2\alpha\beta c_{AP} f^2 I\xi^2 + 2\alpha\beta c_{AL} f^2 I\mu^2\xi^2 - 4\alpha\beta c_{AL} f^2 I\mu\xi^2 + 2\alpha\beta c_{AL} f^2 I\xi^2 - 2\alpha\beta f^2 I\mu^2\xi^2 + 4\alpha\beta f^2 I\mu\xi^2 - 2\alpha\beta f^2 I\xi^2 - 8\alpha\beta f I^2 \mu^2 \gamma + 16\alpha\beta f I^2 \mu\gamma - 8\alpha\beta f I^2 \gamma + 2\alpha f I\mu^2\xi^2\gamma - 4\alpha f I\mu\xi^2\gamma + 2\alpha f I\xi^2\gamma + 4\beta^3 c_{AP}^2 f^2 I^2 \mu^2 - 8\beta^3 c_{AP}^2 f^2 I^2 \mu + 4\beta^3 c_{AP}^2 f^2 I^2 + 8\beta^3 c_{AP} c_{AL} f^2 I^2 \mu^2 - 16\beta^3 c_{AP} c_{AL} f^2 I^2 \mu + 8\beta^3 c_{AP} c_{AL} f^2 I^2 - 8\beta^3 c_{AP} f^2 I^2 \mu^2 + 16\beta^3 c_{AP} f^2 I^2 \mu - 8\beta^3 c_{AP} f^2 I^2 + 4\beta^3 c_{AL}^2 f^2 I^2 \mu^2 - 8\beta^3 c_{AL}^2 f^2 I^2 \mu + 4\beta^3 c_{AL}^2 f^2 I^2 - 8\beta^3 c_{AL} f^2 I^2 \mu^2 + 16\beta^3 c_{AL} f^2 I^2 \mu - 8\beta^3 c_{AL} f^2 I^2 + 4\beta^3 f^2 I^2 \mu^2 - 8\beta^3 f^2 I^2 \mu + 4\beta^3 f^2 I^2 - \beta^2 c_{AP}^2 f^2 I\mu^2\xi^2 + 2\beta^2 c_{AP}^2 f^2 I\mu\xi^2 - \beta^2 c_{AP}^2 f^2 I\xi^2 - 2\beta^2 c_{AP} c_{AL} f^2 I\mu^2\xi^2 + 4\beta^2 c_{AP} c_{AL} f^2 I\mu\xi^2 - 2\beta^2 c_{AP} c_{AL} f^2 I\xi^2 + 2\beta^2 c_{AP} f^2 I\mu^2\xi^2 - 4\beta^2 c_{AP} f^2 I\mu\xi^2 + 2\beta^2 c_{AP} f^2 I\xi^2 + 8\beta^2 c_{AP} f I^2 \mu^2 \gamma - 16\beta^2 c_{AP} f I^2 \mu\gamma + 8\beta^2 c_{AP} f I^2 \gamma - \beta^2 c_{AL}^2 f^2 I\mu^2\xi^2 + 2\beta^2 c_{AL}^2 f^2 I\mu\xi^2 - \beta^2 c_{AL}^2 f^2 I\xi^2 + 2\beta^2 c_{AL} f^2 I\mu^2\xi^2 - 4\beta^2 c_{AL} f^2 I\mu\xi^2 + 2\beta^2 c_{AL} f^2 I\xi^2 + 8\beta^2 c_{AL} f I^2 \mu^2 \gamma - 16\beta^2 c_{AL} f I^2 \mu\gamma + 8\beta^2 c_{AL} f I^2 \gamma - 36c\beta^2 f^3 I^2 \mu^2 + 48c\beta^2 f^3 I^2 \mu - 16c\beta^2 f^3 I^2 - 36\bar{\pi}_{AL}\beta^2 f^2 I^2 \mu^2 + 48\bar{\pi}_{AL}\beta^2 f^2 I^2 \mu - 16\bar{\pi}_{AL}\beta^2 f^2 I^2 - \beta^2 f^2 I\mu^2\xi^2 + 2\beta^2 f^2 I\mu\xi^2 - \beta^2 f^2 I\xi^2 - 8\beta^2 f I^2 \mu^2 \gamma + 16\beta^2 f I^2 \mu\gamma - 8\beta^2 f I^2 \gamma - 2\beta c_{AP} f I\mu^2\xi^2\gamma + 4\beta c_{AP} f I\mu\xi^2\gamma - 2\beta c_{AP} f I\xi^2\gamma - 2\beta c_{AL} f I\mu^2\xi^2\gamma + 4\beta c_{AL} f I\mu\xi^2\gamma - 2\beta c_{AL} f I\xi^2\gamma + 12c\beta f^3 I\mu^2\xi^2 - 20c\beta f^3 I\mu\xi^2 + 8c\beta f^3 I\xi^2 + 12\bar{\pi}_{AL}\beta f^2 I\mu^2\xi^2 - 20\bar{\pi}_{AL}\beta f^2 I\mu\xi^2 + 8\bar{\pi}_{AL}\beta f^2 I\xi^2 + 2\beta f I\mu^2\xi^2\gamma - 4\beta f I\mu\xi^2\gamma + 2\beta f I\xi^2\gamma + 4\beta I^2 \mu^2 \gamma^2 - 8\beta I^2 \mu\gamma^2 + 4\beta I^2 \gamma^2 - cf^3\mu^2\xi^4 + 2cf^3\mu\xi^4 - cf^3\xi^4 - \bar{\pi}_{AL}f^2\mu^2\xi^4 + 2\bar{\pi}_{AL}f^2\mu\xi^4 - \bar{\pi}_{AL}f^2\xi^4 - I\mu^2\xi^2\gamma^2 + 2I\mu\xi^2\gamma^2 - I\xi^2\gamma^2)$

$\sigma_1 = (2\alpha + \beta)/9 - \big(2\alpha(c_{AL} - c_{AP})\big)/9 - \big(2b(c_{AL} - c_{AP})\big)/9 + \big(\beta(c_{AL}^2 - c_{AP}^2)\big)/9 + \alpha^2/9\beta$
$\quad - (2\sigma_2)/(9\beta(3\mu - 2)) + \sigma_2/(9\beta(3\mu - 2)^2) + (2\beta c_{AL} c_{AP})/9$

$\sigma_2 = (\alpha + \beta(1 - c_{AL} - c_{AP})^2$